# A novel approach to preventing SARS-CoV-2 transmission in classrooms: An OpenFOAM based CFD Study

Anish Pal[1], Riddhideep Biswas[1], Ritam Pal[2], Sourav Sarkar[1, *], Achintya Mukhopadhyay[1]

[1]Department of Mechanical Engineering, Jadavpur University, Kolkata-700032, India

[2]Department of Mechanical Engineering, The Pennsylvania State University, University Park, Pennsylvania 16802, U.S.A.

*Corresponding author: souravsarkar.mech@jadavpuruniversity.in

**Abstract**

The education sector has suffered a catastrophic setback due to ongoing COVID-pandemic, with classrooms being closed indefinitely. The current study aims to solve the existing dilemma by examining COVID transmission inside a classroom and providing long-term sustainable solutions. In this work, a standard 5m x 3m x 5m classroom is considered where 24 students are seated, accompanied by a teacher. A computational fluid dynamics simulation based on OpenFOAM is performed using a Eulerian-Lagrangian framework. Based on the stochastic dose response framework, we have evaluated the infection risk in the classroom for two distinct cases: (i) certain students are infected (ii) the teacher is infected. If the teacher is infected, the probability of infection could reach 100% for certain students. When certain students are infected, the maximum infection risk for a susceptible person reaches 30%. The commonly used cloth mask proves to be ineffective in providing protection against infection transmission reducing the maximum infection probability by approximately 26% only. Another commonly used solution in the form of shields installed on desks have also failed to provide adequate protection against infection reducing the infection risk only by 50%. Furthermore, the shields serves as a source of fomite mode of infection. Screens suspended from the ceiling, which entrap droplets, have been proposed as a novel solution that reduces the infection risk by 90% and 95% compared to the no screen scenario besides being completely devoid of fomite infection mode. As a result of the screens, the class-time can be extended by 55 minutes.

## 1. Introduction

The evaporation and dispersion of respiratory droplets have been studied extensively after the onset of the Covid-19 pandemic in 2020 as they played a major role in the spread of the virus[1,2]. It has been established that the trajectory of the droplets is influenced by the droplet evaporation. The spreading of droplets depends on the surrounding conditions such as temperature, relative humidity and wind speed. This implies that the indoor and outdoor conditions will produce different droplet trajectory even if the initial velocity of the droplets is same. An extensive amount of numerical and analytical work has been carried out to investigate the spreading of respiratory droplets in an outdoor environment. The effect of wind-speed and direction, ambient pressure and relative humidity on respiratory droplets was extensively studied as well as detailed studies pertaining to the type of respiratory activities and the concomitant risk has been reported [3–8]. All studies indicated that maintaining proper social distances can significantly reduce the infection probability.

The transmission of droplets in an indoor environment is very much different from the outdoor environment because the ventilation in an indoor environment plays a crucial role in the trajectory of the droplets. Improper ventilation in an indoor environment can cause droplets to linger in the domain for extended periods. Furthermore, maintaining proper social distance in an indoor setting is not always a possible strategy of minimizing infection risk. Considering the extra risk associated with indoor settings researchers have investigated various indoor settings like airplane cabins, buses, elevators and courtrooms [9–13]. Cheng et al. analytically investigated the trajectories of large respiratory droplets in indoor environment[14]. The droplets with diameters 40-200 micron were studied in this work and the travel distance of these droplets were 0.16-0.68 m for speaking, 0.58-1.09 m for coughing and 1.34-2.76 m for sneezing. Pal et al. performed an analytical study to report the effect of atmospheric conditions on the evaporation and transport of respiratory droplets in indoor environment[15]. The work considered a salt solution droplet as a respiratory droplet to make it analogous to a real pathogen laden respiratory droplet. The study showed that complete evaporation of droplet lead to the formation of aerosols that remain suspended in the air while carrying a significant amount of viral load. The article provided with a risk map corresponding to different weather conditions. It showed that hot and dry conditions were most favourable to aerosol formation and the aerosols were carried up to distances of 8 to 9 m. Li et al. conducted an experimental study to evaluate the transmission characteristics of respiratory droplets and aerosol in indoor environment[16]. The results showed in the presence of ventilation, the viable droplets were found to be less than 3.3 micron in size but without any air change, the viable droplets were larger than 3.3 micron in diameter. Gao et al. adopted Eulerian modelling approach to study the distribution of respiratory droplets in enclosed environments[17]. The study included the different ventilation methods such as mixing ventilation, displacement ventilation, and under floor air distribution. The results showed that the respiratory droplets got trapped in the breathing height of people in displacement ventilation, and under floor air distribution. Mixed ventilation resulted in a uniform distribution of droplets. So, low air speed in the breathing area could lead to prolonged human exposure to infectious droplets. Biswas et al. performed a computational analysis to investigate the trajectory of coughed droplets in an elevator under different ventilation conditions[18]. A risk factor was evaluated in the study based on the fraction of droplets lingering in the mouth to waist zone. The article reported that the presence of exhaust fan at the elevator ceiling produced the safest condition with a risk factor of zero while the quiescent environment was the most unsafe environment. Pal et al. numerically investigated the airborne and fomite risk based on a stochastic-dose response model in an elevator under various ambient conditions and determined the appropriate fan speed for ensuring minimum risk[19]. Ng et al. reported the growth of respiratory droplets in a cold humid ambience[20]. Sen studied the transmission and evaporation of cough droplets in an elevator considering a top mounted fan at the ceiling and bottom outflow slots at the walls[21]. The absence of the fan created a risky situation in the elevator with the droplets hanging around but when the fan was switched on, the droplets settled on the ground. Dbouk et al. investigated airborne virus transmission in elevator and confined spaces[22]. Different flow scenarios were considered in this study by varying the inlets and outlets of the air and by including the operation of an air purifier. The study concluded that the presence of air purifier was not helpful in preventing virus transmission and the presence of single inlet and single outlet exhibited the least droplet dispersion. Rijn et al. performed some experiments to find out a way to reduce the transmission of SARS-Cov-2 inside hospital elevators[23]. A typical 10-20% open door time showed that it took 12-18 minutes for the number of aerosol particles to decrease 100-fold in both medium- and large-sized elevators.

The education system has hit a huge road-block during this pandemic and has not yet returned to normalcy across the globe. Efforts are being undertaken to bring back the education system to pre-pandemic mode. However, in numerous cases the reopening of schools has followed a wide-spread infection among teachers and students[24] which in many cases has resulted closing of schools again. This continuous interruption towards complete return to normalcy education needs to removed. To ensure smooth reopening of the education systems and sustain it, ways to reduce the covid-transmission risk within a classroom needs to be devised. A limited number of studies have been conducted on covid-transmission in an indoor setting like classroom, although extensive studies have been carried out in other indoor settings as discussed above. Quinones et al. reported the evolution of size of respiratory droplets in a classroom and this paved the way for some preventive measures[25].

Chen et al. studied the effect of ventilation and different source locations in a classroom on the spread of respiratory droplets by experimenting with low-cost sensors[26]. They developed a particle concentration monitoring network in a classroom to explore the dispersion of droplets. The results showed that the instructor or the source should be located opposite to the ventilation flow, and the air handling unit and the fan coil unit must be turned on during class hours irrespective of the outside weather conditions. Burgmann et al. conducted both experimental and numerical evaluation to reduce the transmission of aerosol in classrooms[27]. They employed an air-purifier system and it was observed that the air-purifier led to a significant reduction of airborne particles in the room depending on the infected person's location. Mirzaei et al. reported infection risk reduction through implementation of partitions around students[28]. Arjmandi et al. worked on the optimization of ventilation systems in a classroom with the help of numerical modeling[29]. Certain studies reported the implementation of personal ventilation system to minimize the risk transmission[30–32].

The limited number of studies have failed to address certain very important aspects associated with risk transmission in a classroom. In this current study we have tried to address these aspects which have enumerated below.

- There has been lack of literature data on the risk transmission in a classroom setting considering the student is infected. Majority of the studies have considered the teacher is infected. In the current study both the scenarios of the student and teacher being infected has been considered and investigated in detail. The ambience has been selected as the hot dry quiescent ambience (30°C, 30 % R.H.) as this ambience pose the highest infection risk[19]. Furthermore, the maximum infection risk that cumulates in the classroom depends upon the position of the infected student. Hence, in this study the most risk posing locations has been identified through a rigorous process and the students in that locations have been assumed to be infected. Designing apposite strategies for reducing infection transmission associated with these infected students seated in these most precarious positions will be applicable to all other scenarios. The shortcomings associated with commonly used preventive measures has been identified and a novel and low-cost sustainable solution in the form of transparent polycarbonate screens suspended from the classroom ceiling has been proposed along and has been shown to be efficacious in preventing both airborne and transmission risk.

- Majority of students have considered the respiratory droplets to be composed of pure water which does not represent a real-world setting. Non-volatile salts in precise quantities are present in the droplets, and these salts carry pathogens. A salt-laden droplet will have different thermophysical characteristics than a pure water droplet, and the thermophysical qualities will vary with evaporation, unlike pure water. This discrepancy and continual change in thermophysical characteristics will result in a difference in the mass transfer number and, consequently, the evaporation rate, resulting in a difference in total droplet dispersion and trajectory compared to pure water. Hence, the breath droplets have been constituted of salt solution in this study to represent the real-world scenario.

- Majority of studies have considered the wells-riley model for evaluating the infection-risk in a classroom[33] or for an indoor setting [34–36]. However, there are some key downsides of the Wells-Riley model. To begin, the Wells-Riley model is based on the premise that the indoor space under examination is a well-mixed environment[34]. As a result, in scenarios where ventilations are either unable to provide a well-mixed atmosphere or have a directed character, this assumption is invalid. Second, for any new ventilation condition, the Wells-Riley model requires backward derived quanta generation rate data, which is not accessible[37]. Third, the Wells-Riley model involves various implicit mistakes (such as geometry, ventilation, and infectious source strength) through the backward computed quanta production rate, and hence the model cannot be employed properly in any general case[37]. Hence, for all these disadvantages associated with the wells-riley model, the dose response model has been implemented to follow a more authentic approach for risk formulation. Furthermore, Liu et al. analysed four risk assessment models, including the exposure risk index, intake fraction, improved Wells-Riley model, and dose-response model, and determined that the dose-response model is the most accurate[38].

- Majority of studies have investigated the airborne mode of infection in a classroom. In a classroom susceptible persons stay for an extended time and during their extended stay they might touch various surfaces. Virus containing droplets settling on these surfaces provide the fomite mode of infection pathway. Hence, the investigation of the fomite mode of infection in the classroom needs to be investigated. This study not only investigates the fomite mode of infection but also reports that the fomite mode of infection constitutes a significant infection-risk.

- As droplets are ejected into the domain continuously due to the various respiratory activities the infection-risk increases. Hence, it is important that the maximum class-time that maintains sufficiently low infection risk in the domain be determined. This study besides investigating the risk propagation in the domain in details predicts the maximum allowable class-duration.

- The droplet size has severe epidemiological implications. Based on the droplet size the virus containing droplets gets deposited on various tracts of the respiratory tract, i.e., the alveolar and bronchial (lower respiratory tract) or the Extrathoracic region (upper respiratory tract). Infection in the lower part of the respiratory tract is more detrimental than in the upper part of the respiratory tract[39]. Furthermore, the small-size droplets can deposit in the olfactory epithelium and infection in the olfactory epithelium can cause to olfactory dysfunction and anosmia[40]. Infection in the olfactory epithelium can also lead to neurological implications and long-term damage to brain[41]. Hence the understanding of epidemiological implications associated with droplet size is very important and, in this study, these epidemiological aspects of the virus containing droplets has been investigated in this study.

In the present study, the classroom has been considered to be composed of 24 students and 1 teacher[42–44]. Transmission of droplets emanating from both the teacher and the students has been considered. In this work a three-dimensional simulation has been performed to understand the transmission of virus laden breath generated droplets in a classroom. Computational Fluid Dynamics (CFD) approach using a Eulerian-Lagrangian framework, utilizing opensource software OpenFOAM has been employed to simulate the above situation. The droplets are considered to be composed of a saline solution containing 1% NaCl (salt) and 99% $H_2O$ by wt[45]. The risk of infection is quantified by a parameter called risk factor, whose formulation is based on the stochastic dose-responsive model[46]. This study proposes a low-cost sustainable solution for preventing SARS-CoV-2 transmission inside a class room. The infection risk for each susceptible person in the classroom, has been enumerated for various covid avoiding strategies.

## 2. Problem Formulation

### I. Geometry

The domain under consideration is a 5m X 5m X 3m classroom in which 24 students and one teacher are seated[42–44] The students having a seating height of 1m[47] are all seated on individual benches, 0.4m X 0.35m X 0.85 and desks having height 0.55m in 4 rows. The rows in the left half and right half are staggered by a distance of 0.2. The class room constitute two windows each 1.5m X 1m and a door, 2.1m X 1m. Figure. 1 depicts the all the constituents of the classroom along with the location of various students (students are numbered for easy identification) and the teacher. The students have been numbered for ease of identification. The features of the teacher and students has been considered as rectangular features for the reduction of computational cost and complexity. The mouth from where the droplets are continuously ejected has been considered to be of a rectangular slot of 40 mm x 5 mm[48].

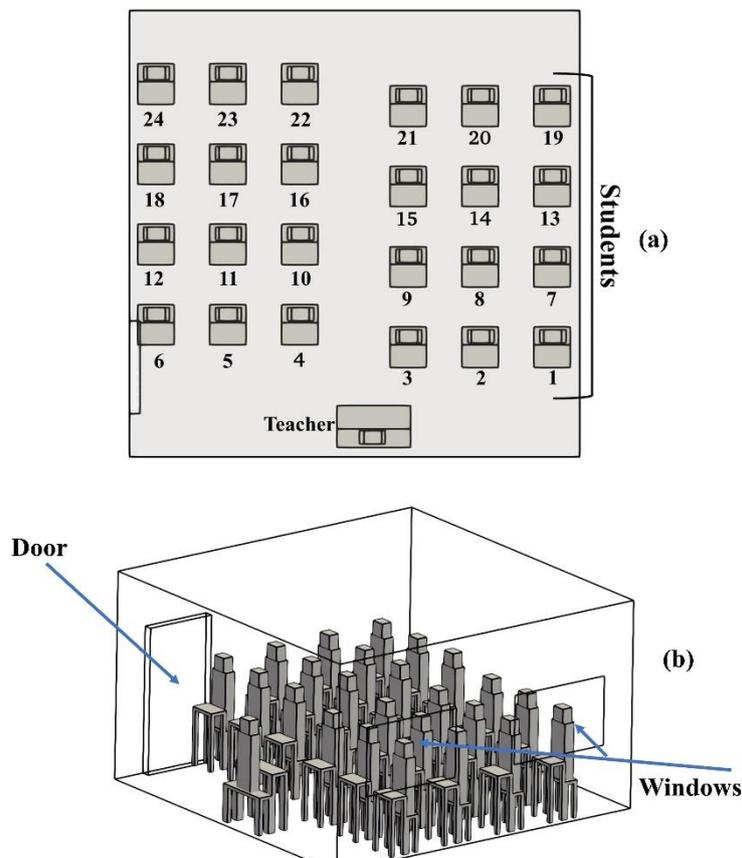

**Figure 1(a)-(b): Showing the location of the teacher, students ( students are numbered) , door and windows within the computational domain.**

### II. Initial and Boundary Conditions

The walls and floor of the room, the benches and desks, the surfaces of students and teacher are considered as walls with a no-slip velocity. The doors and windows are assumed to be open (and hence modelled as pressure outlet), thus maintaining natural ventilation in the classroom. The ambient condition is considered to be hot dry, 30°C, 30 % R.H.  Any afflicted person, has a 38.4°C body temperature, whereas the body temperature of the susceptible persons is assumed to be 37°C. Continuous inhalation and exhalation of all the persons following a sinusoidal profile has been considered [49]. The exhalation air expelled from the mouth is considered to be at body temperature, with a relative humidity of 100%[50]. Droplet size pertaining to the breathing droplet size has been considered to be emanating from students and for the teacher the speaking droplet size has also been additionally incorporated[51,52] as depicted in Table 1. A class-duration of 100s has been taken into consideration, since the infection risk cumulates to a significantly high value (In some cases infection probability reaches 100%) within this time-frame. The aim of the study is to solutions which completely eradicates the risk. Hence, if apt strategies are developed which can completely eradicate risk within the time-period of 100s (within which there is a significant risk-build-up), such solutions can continue to maintain a low-risk situation for extended periods.

| Table 1: Breathing droplet Size ||| 
|---|---|---|
| **Sr. No.** | **Size(μm)** | **Particle/second** |
| 1 | 0.3-0.5 | 87 |
| 2 | 0.5-1 | 46 |
| 3 | 1-3 | 22 |
| 4 | 3 | 2 |

**III. Mathematical Modelling**

The authors' previously constructed numerical model was employed in this investigation[53]. A fully coupled Eulerian-Lagrangian model was employed for this numerical analysis. To simulate the carrier fluid, air, the Eulerian frame is utilised. For the carrier bulk multiphase continuity and momentum equations in conjunction with the k - SST turbulence model(s) were employed[53,54]. In the Lagrangian reference, the injected respiratory droplets are solved as discrete particles[53,54]. The droplet position and velocity are calculated by taking gravity, buoyancy, lift, and drag forces. The Nusselt number and the Sherwood number were calculated using the Ranz-Marshall model[55,56]. The temperature of the droplet is determined by the energy conservation equation[53,54]. Cough droplets are represented as a combination of water and NaCl (99% water and 1% NaCl by weight)[45]. When droplets evaporate and lose their volatile liquid mass to the environment, the mass fractions of the components change. Droplet characteristics are determined by the properties of liquid water and salt, as well as their instantaneous mass fractions. Droplet evaporation causes a constant change in the mass fractions of the constituents. Changes were made to the source code of OpenFOAM's reactingParcelFoam[54] solver to accommodate this salt model of droplets. The precise equations are included in the paper's supplemental section for the convenience of any interested reader. The OpenFOAM (an open-source CFD-solver) solver "reactingParcelFoam" was used to solve the partial differential equations, with the necessary alterations to properly apply the model for the salt solution as previously described. It is critical to note that the thermophysical properties of both the Lagrangian and Eulerian phases are temperature dependent. The Eulerian phase has been classified as an ideal gas for its equation of state, and its transport has been simulated using Sutherland's law[57] of viscosity, which is based on the kinetic theory of gases and applies to non-reacting gases. The Eulerian phase was discretized using finite volume techniques. Schemes of the second order are utilised for both space and time operators. In our current work, we are striving to make our simulations more realistic by including the salt model of droplets, which accounts for the effect of salt solution in cough droplets. Our salt model closely matches the experimental data of Basu et al.[45] The overall numerical model including the carrier phase dynamics as well as the droplet transport has been validated against the DNS data of Ng et al.[20] These comparisons are detailed and quantified in the authors' earlier work[58]. In order to increase paper readability, a comprehensive grid-independent and time-independent investigation was conducted and presented in the appendix section.

**3.Results and Discussion**

In this study risk propagation within a classroom subjected to a hot dry ambience (30ºC, 30 % R.H.) has been evaluated, since this ambience has been proven to most conducive for risk propagation[53] . Furthermore, devising apposite strategies for risk minimization in this climatic ambience will suffice all other ambiences. The risk propagation has been considered for two different scenarios: (i) certain students are affected (ii) the teacher is effected has been evaluated. The risk formulation is based on the stochastic dose response model. A breathing box, each of size 0.3m x 0.4m x 0.3m, has been considered near each person's mouth as depicted in Fig. 2. The Risk of Infection (R) is calculated in all of these boxes to evaluate the infection risk of each person using a dose-response model as per equation 1 below.

$$R = 1 - \exp(-\sigma\mu) \qquad (1)$$

μ is the expected number of pathogens likely to be inhaled by a susceptible person over the exposure time of 100s given by equation 2.

$$\mu = \sum_\beta \int_0^1 \frac{P}{V} \frac{\int_0^{T_0} \iiint_V \frac{c\pi}{6} D_0^3 N \, dt}{T_0} dt \qquad (2)$$

where $D_0$ is the initial diameter of injected cough droplet; N is the mean viral load in the respiratory fluid of covid-19 infected persons (7 X 10$^6$ RNA copies/ ml)[59]. P is the pulmonary ventilation rate, V is the volume of the considered breathing box and β is the total number of inhalation cycles over the total exposure time ($T_0$), considering an inhalation period of 1s. $\frac{P}{V}$ represents the probability of inhaling a droplet suspended within the breathing box. The infectivity factor, "σ" is the inverse of the number of viruses capable of initiating infection and is a direct indication of the infectious dosage[46]. Infectivity factor for SARS-COV-2 as reported by a recent study by Mikszewski et al[60] has been used for the current work.

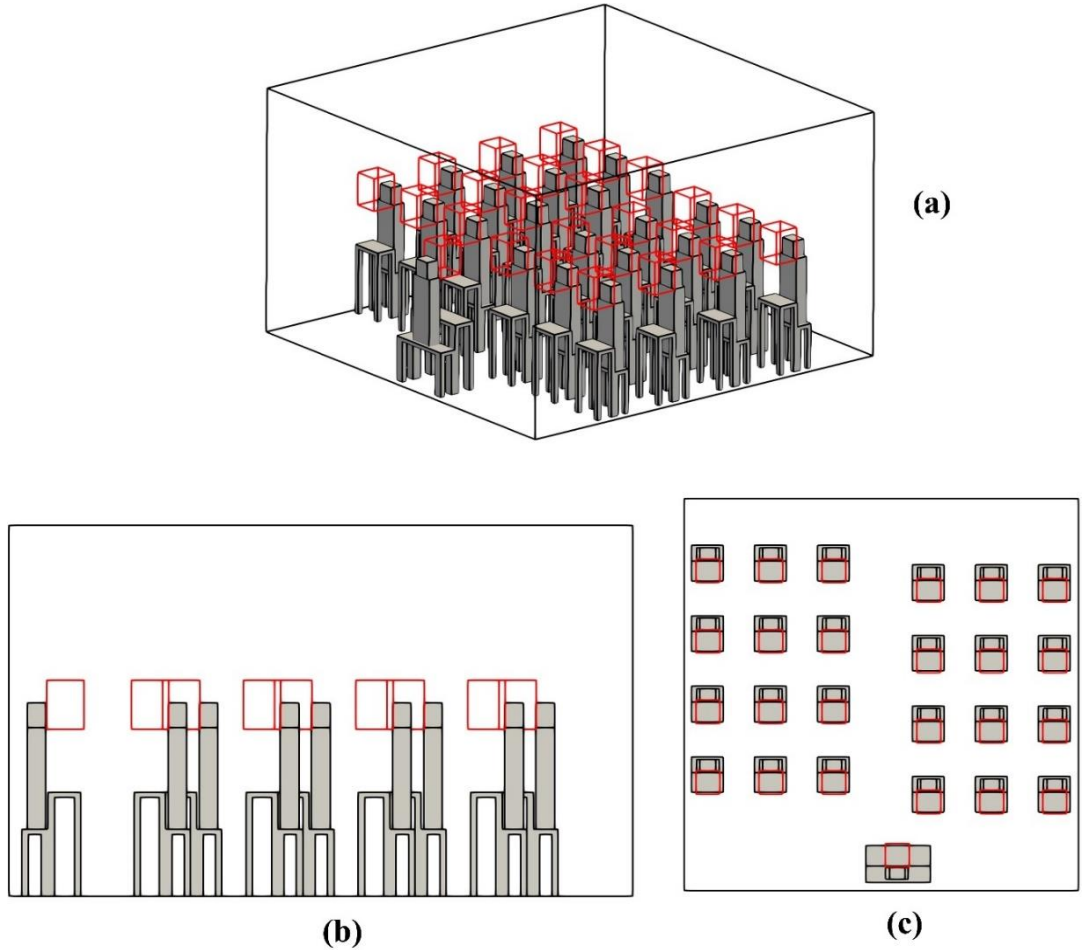

**Figure 2 (a)-(c): Showing isometric, sideview and top-view of the breathing boxes in the vicinity of the students**

**3.1. Transport and epidemiological implications of droplet emanating from infected students and teacher**

**I.A. Quiescent Scenario (considering students are infected**

Initially the infection risk of the susceptible persons based on a stochastic dose response model, considering only one student has been infected was computed. This analysis was carried out separately for each of the 24 students. The risk heat map associated with that study which depicts the risk transmitted by each individual student to all other susceptible students and the teacher, has been reported in Fig.1 the appendix I section. The total cumulative risk posed by a posed by a single student (equation 3) has been depicted in Fig.3 for each individual student.

$$R_i^{total} = \sum_{j=1}^{24} R_{i-j} \, (i \neq j) + R_{i-Teacher} \quad (3)$$

j represents the students and hence has been varied from 1 to 24., $R_{i-j}$ represents the risk from *ith* student to *jth* student and $R_{i-Teacher}$ represents the risk from ith student to the teacher.

Figure 3 demonstrates that students numbered (student numbering has been depicted in details in Fig.1) the students seated inline with the doors and windows pose high risks whereas those away from any door and window pose very low risk. As discussed earlier, the ambient condition has been chosen as a hot dry condition (30°C, 30 % R.H.) in a quiescent ventilation scenario with the doors and windows being open and providing natural ventilation. The droplets injected into the domain due to the inhalation and exhalation of the students being in the aerosol size range[51,52] move upwards upon injection. In addition to the buoyancy and lift force causing the vertical upward movement of the droplets, the natural convection caused due to the thermal stratification produced due to body temperature[61] of the students also play an important role in the upward movement of the droplets. Besides the vertical natural convection in the domain, a lateral convection is also generated near the doors and windows as demonstrated in the velocity vector plots of Figs.4-5. However, in contrary to this understanding it has been observed that the students just adjacent to the doors and windows pose significantly lower risk as compared to those seated in line but not adjacent to windows and doors. This is because the droplets emanating from these students get entrapped in the sidewalls thus impeding the lateral spread of the droplets which ultimately causes the risk posed by these students. The droplets from sources seated away from windows and doors mainly undergo a vertical movement causing them to linger above the breathing box's height and ultimately reducing the infection risk. A comparison of the velocity vectors juxtaposed with droplet dispersal in the domain depicted in Figs.4 and 5 for student 4 and student 24 respectively further elucidates the issue. The student numbered 4 is seated inline with the door whereas the student number 24 is away from all doors and windows. The effect of lateral convection on droplet dispersal an its corresponding lateral dispersion is evident in Fig.4. However, as can be seen from Fig.5 for student numbered 24 the absence of any kind of lateral convection renders the infection transmission from such a student positioned far away from the windows and doors to a very low magnitude. Thus, we understand from the above discussion that lateral convection has a very important role on the droplet dispersal and hence ultimately on risk-transmission in the domain. The total risk pertaining to each student as depicted in Fig.3 serves as a testament to this argument. The students numbered 4,14,15,20 pose the highest risk in the domain due to reasons discussed above. From, hereon the droplet dispersal of droplets emanating from students 4,14,15,20 and the associated risk has been studied and represented.

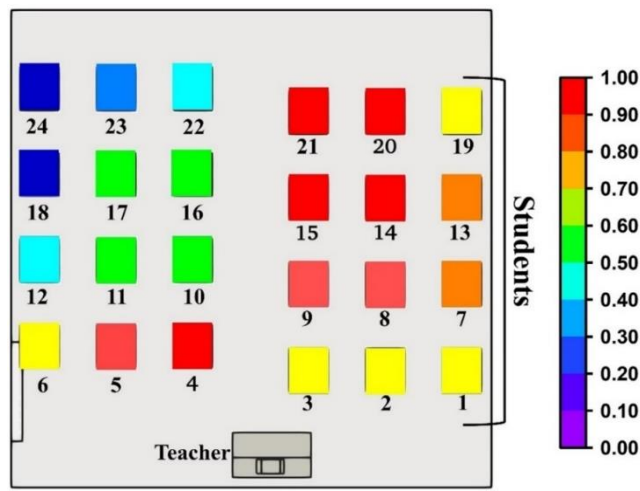

**Figure 3: Depicting the total risk posed by each student to all other susceptible persons in the classroom**

Figure 6 reports the infection probability for each susceptible student and the teacher (depicted by the symbol T) posed by the above identified four infected students evaluated over an exposure period of 100s based on a stochastic-dose response model (equation 1-2). It depicts over an exposure period of 100s the infection risk reaches a significantly high magnitude of 40% in the quiescent scenario with natural ventilation of windows and door. It can be concluded from Fig.6 that basic reproduction number($R_0$)[62] of the classroom reaches a very high magnitude of 6 thus creating a very grave situation. The susceptible persons adjacent to the infected persons and seated inline are the most affected persons thus emphasizing the effect of lateral convection on droplet dispersal and thus risk transmission. From, hereon the droplet dispersal of droplets emanating from students 4,14,15,20 and the associated risk has been studied and represented.

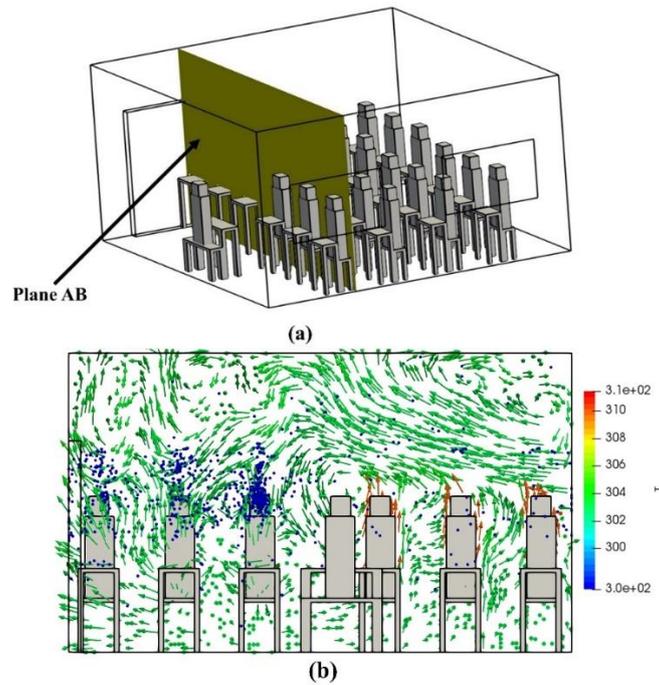

**Figure 4(a): Showing Plane AB where the velocity vector are plotted**
**4(b): Showing temperature contoured velocity vector plot showing the lateral convection causing the lateral spread of droplets( on Plane AB)**

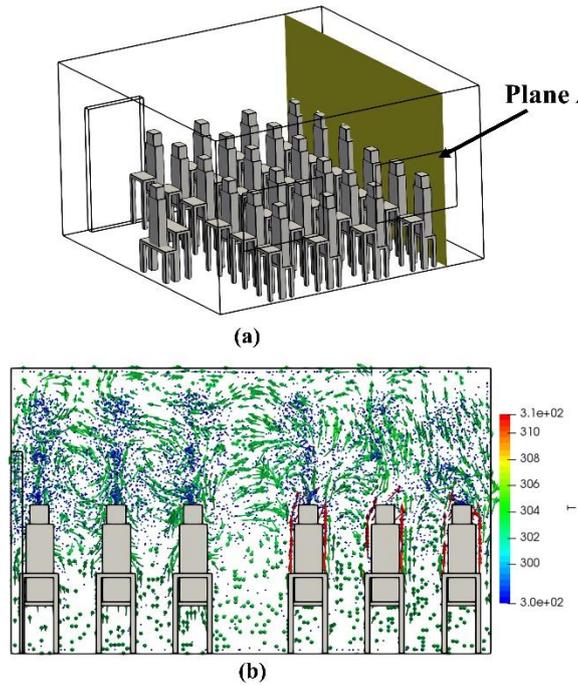

(a)

(b)

**Figure 5(a): Showing Plane AB where the velocity vector are plotted
4(b): Showing temperature contoured velocity vector plot showing lack of lateral convection(on Plane AB)**

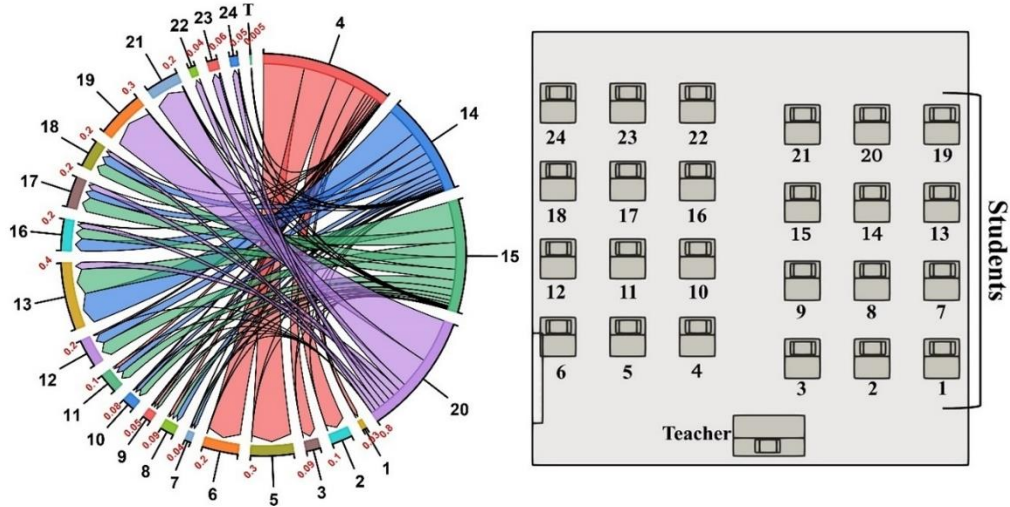

**Figure 6: Quantifying the airborne-risk transmission from the infected to susceptible persons(left) and the location of the students and teacher(right).**

**I.B. Quiescent Scenario considering the teacher is infected.**

The previous discussion up till was pertaining to the infection risk transmission concomitant with the infected students. However, another aspect that needs to be investigated is the risk transmission in the classroom considering the teacher is infected.

As discussed earlier the mask usage leads to excessive vocal strain, vocal effort, elevated speech level and ultimately leads to vocal fatigue[63]. Furthermore, the study suffices to all socio-economic settings that includes classrooms that are devoid of any kind of sound-relaying system. Hence, to ensure an extended class where the voice of teacher is audible to each student without causing major discomfort of teacher it has been assumed that teacher is not wearing any mask. Droplets pertaining to corresponding sizes injected during both speaking and breathing have been considered[51,52]. Analysis reveals in a quiescent medium the teacher poses significant risk, the infection probability reaches 100% (magnitude of Risk is 1) for certain students thus creating a grave situation. Figure 7 depicts in detail the risk teacher poses to each student. The infected teacher poses significant infection risk to 6 students as can be understood from Fig.7. In the case of teacher being infected, the infection risk is not only limited to airborne risk but also entails a significant quantity of fomite risk. While the smaller size droplets ejected during breathing linger in the domain and contribute to airborne risk as depicted in the droplet dispersion Figs 8-9, a significant percentage of larger droplets emanating from the mouth during speaking settles down on the desks of the students thus creating a fomite infection pathway. The Nicas and Best model was used to assess the risk of fomite infection on the desks[64]. Because of the significantly less decay rate on the desk surfaces (assumed to be wood) Nicas and Best[64] model has been implemented (equations 3-5) using relevant literature data[64–66].

$$E_m = f_m c_m A_s \overline{C_{\text{hand},t_o}} t_o \tag{3}$$

$$\overline{C_{\text{hand},t_o}} = \frac{f_b c_h C_s}{(\varphi + f_h c_h + f_m c_m) t_0}\left[t_0 + \frac{\exp(-(\varphi + f_h c_h + f_m c_m) t_o) - 1}{\varphi + f_h c_h + f_m c_m}\right] \tag{4}$$

$$R = 1 - \exp(-\sigma E_m) \tag{5}$$

$E_m$ being the dose of pathogen delivered to the mucous membrane, $c_h$ is the pathogen transfer efficiency from the surface to the hand after a contact, $c_m$ is the pathogen transfer efficiency from the hand to the mucous membrane after a contact, $f_h$ is the frequency of hand-to-contaminated surface contact, $f_m$ is the frequency of hand-to-mucous membrane contact, $A_S$ is the average contaminated surface area touched per hand contact, $t_o$ is the concerned time interval, $C_s$ is the pathogen load per area of the contaminated surface, and $\varphi$ is the decay rate of the pathogen on hand. The fomite risk depicted in Fig.10 demonstrates a significant fomite infection risk exists.

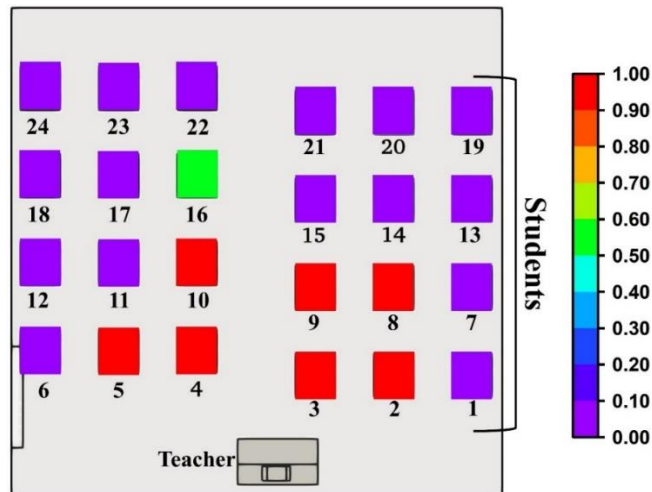

**Figure 7: Airborne-infection risk posed by the infected teacher to the susceptible students.**

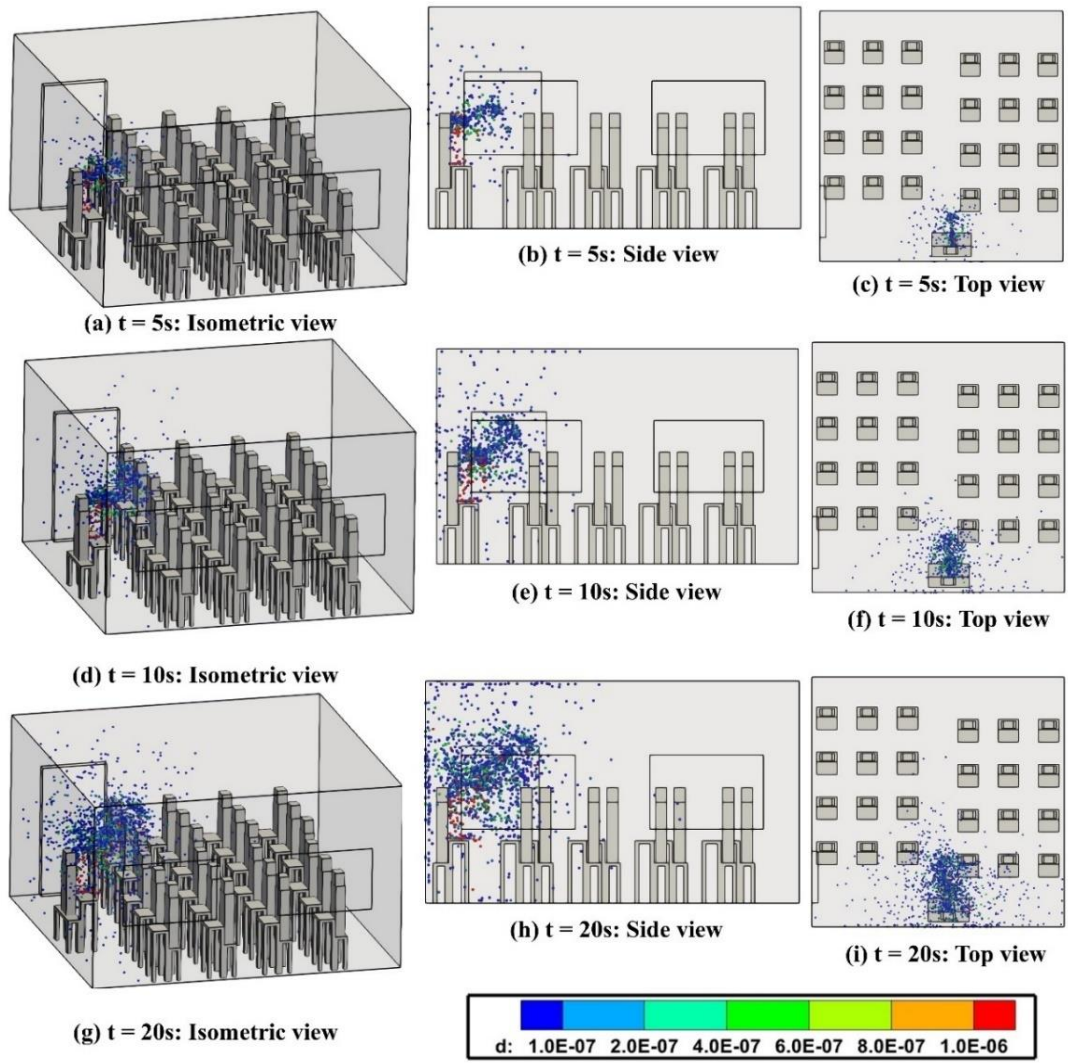

**Figure 8(a)-(i): Dispersal of droplets ejected by the teacher ( up to 20s)**

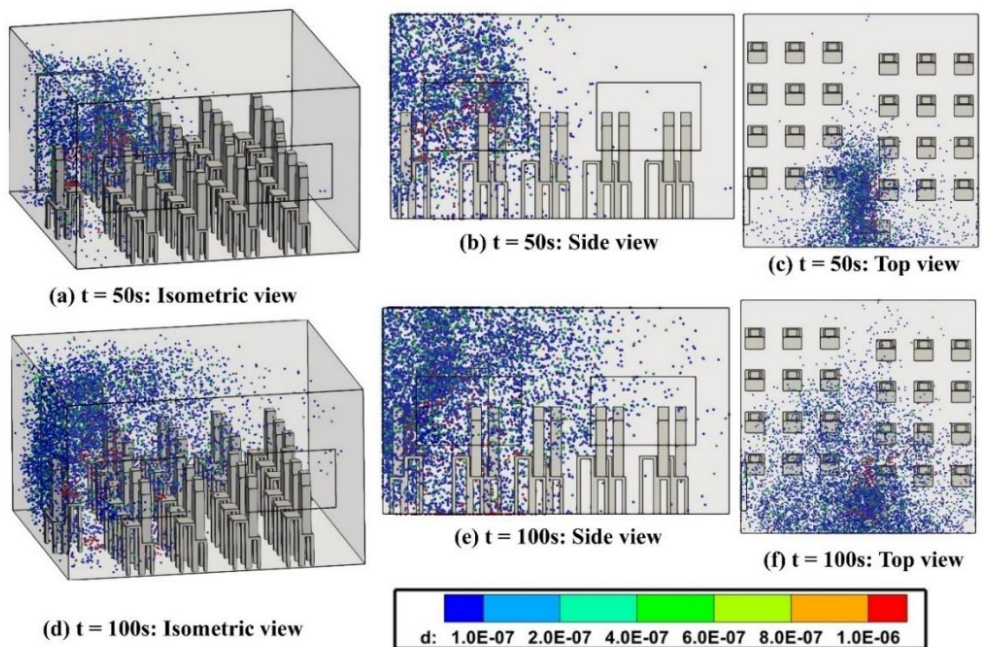

**Figure 9(a)-(f): Dispersal of droplets ejected by the teacher ( up to 100s)**

**Submitted to Arxiv**

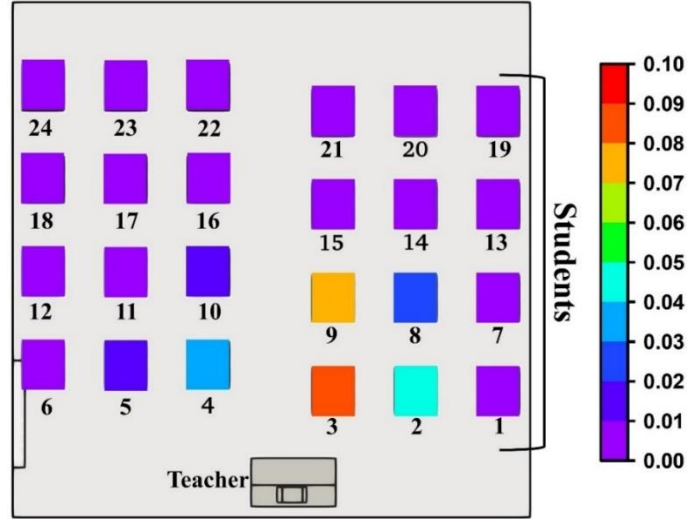

**Figure 10: Plot showing fomite risk posed by the infected teacher.**

**I.C. Epidemiological ramifications of the droplet size in a Quiescent domain.**

The risk formulation based on the dose response model dictates that the infection probability depends upon the virion quantity inhaled, however it does not take into account the droplet size inhaled. However, the droplet size play a major a role on infection severity and its ramifications. Droplet sizes greatly affects in which part of the respiratory tract they end up , Extrathoracic (upper Respiratory tract) or Alveolar and Bronchial ( lower Respiratory Tract) [67] or in the olfactory epithelium[40,68]. The droplets ejected during breathing are mainly sub-5μm droplets, hence the ejected droplets or the suspended evaporated droplets or droplet nuclei get deposited mainly in alveolar and bronchial regions. Deposition in the lower parts of the respiratory tract, namely the bronchial and alveolar is a alarming issue, since deposition in these areas can lead to long term lung damage as well as it may require critical diagnosis. Virion deposition in the olfactory epithelium leads to olfactory dysfunction and anosmia[40] . Owing to the size range of the suspended droplets in domain, as portrayed in the droplet size distributions depicted in the droplet dispersion figures above, a significant percentage of these virus containing droplets get deposited within the olfactory epithelium. This leads to olfactory dysfunction and is associated with focal opacification and inflammation of the olfactory epithelium[40]. This phenomenon is a very grim situation  because this not only leads to impaired olfaction but also as recent studies have shown SARS-CoV-2 can infiltrate the neurological system by crossing the neural-mucosal interface in the olfactory mucosa and taking advantage of the close proximity of olfactory mucosal, endothelial, and neuronal tissue, including sensitive olfactory and sensory nerve endings[69] The probable particle deposition count in the bronchial and alveolar region $D_{c_{BA}}$ and in the olfactory epithelium $D_{c_{Olf}}$ within any breathing, box is quantified through a formulation as per equation 4.

$$D_{c_{Olf}} = \langle \Sigma P_{i_{Olf}} N_{D_i} \rangle_{B_j} , D_{c_{BA}} = \langle \Sigma P_{i_{BA}} N_{D_i} \rangle_{B_j} \qquad (4)$$

;where $N_{D_i}$ is the total number of particles of diameter $D_i$ suspended within the breathing box $B_j$ and $P_{i_{Olf}}$ and $P_{i_{BA}}$ are the deposition efficiencies[68] in the olfactory epithelium and bronchiolar and alveolar regions for that size of droplet respectively. The respective temporally-averaged count of droplets within a breathing box that gets deposited on the olfactory epithelium and the alveolar and Bronchial regions, are depicted in Figs.11-12 for the quiescent hot dry scenario for both the scenarios of teacher and student being infected respectively.

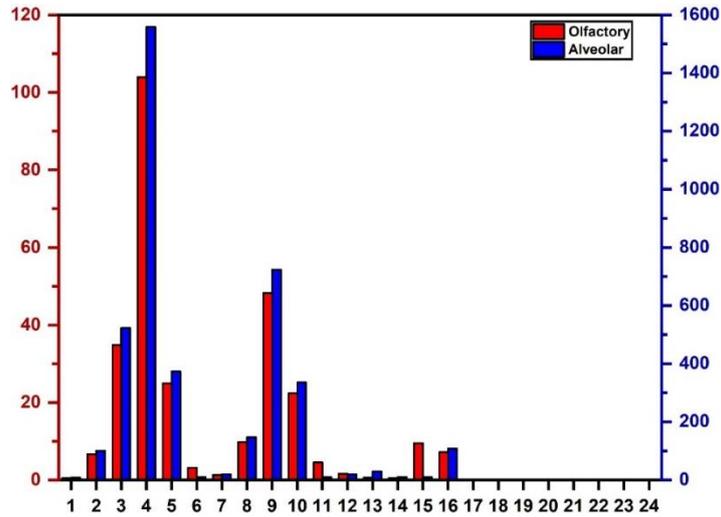
**Figure 11: Depicting the droplet deposition count on in the alveolar and olfactory epithelium for the teacher infected scenario.**

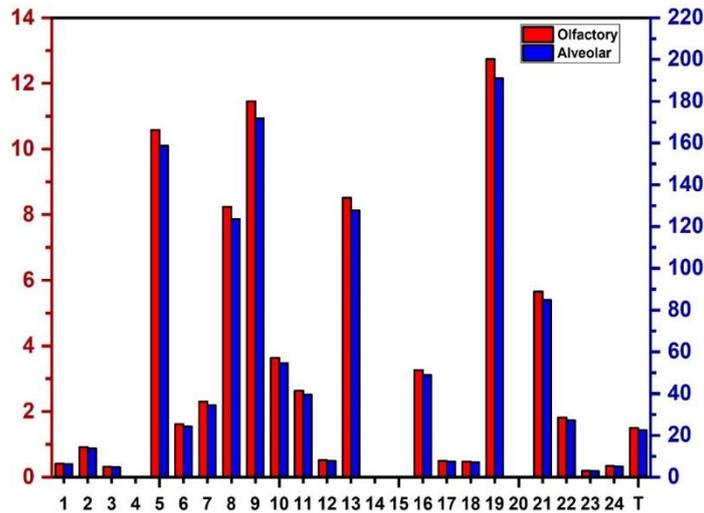
**Figure 12: Depicting the droplet deposition count on in the alveolar and olfactory epithelium for the student infected scenario on susceptible persons (*4,14,15,20 are themselves infected hence no droplet count has been shown) .**

Thus we observe that not only the quiescent scenario poses significant infection risk but also the epidemiological ramifications are quite severe thus strategies need to be devised to reduce the infection transmission. Thus the following sections of this paper explores the various risk-reducing strategies. Besides demonstrating the various draw-backs associated with each strategy, a novel solution has also been proposed that eliminates all the draw backs of other strategies and reduces the infection probability significantly.

**II. Risk reduction through mask usage**

The previous section describes that in a quiescent scenario with the natural ventilation of windows and door creates a significant perilous situation in the domain within an exposure time of 100s. Hence, a low-cost sustainable solution for ensuring a safe domain is the usage of masks by all the students. The fairly ubiquitous and economically affordable to all socio-economic classes of the society cloth mask has been considered in this study to evaluate in providing protection against infection. Face masks, regardless of usage, enhance the feeling of vocal strain, difficulties in word recognition and trouble harmonizing speaking and breathing. Individuals who used face masks for professional and vital tasks reported more symptoms of vocal pain and exhaustion, vocal strain, difficulty with speech perception, and troubles coordinating speaking and breathing[63]. Furthermore, it has been considered that the classroom is devoid of any sound system that relays the voice of the teacher. The study has been undertaken such that the results are applicable in all socio-economic conditions and classrooms in many developing nations are devoid of such relaying sound systems. To ensure that the teacher can take an extended class without discomfort, fatigue by talking in a manner that does not require speaking at elevated levels the teacher has been considered not wearing a mask. The change in droplet trajectories the hindrance caused by the mask has not been taken into account considering the particle size[51,52] in consideration[70]. The formulations pertaining to masks are based only on the droplet being injected upon the filtration efficiency of the corresponding droplet size[71] as well as considering similar inhalation and exhalation mask efficiency[72]. There has been recommendation from the world health organization[73], various state and national governing bodies[74] and some studies[75] towards the usage of cloth mask by the general public. A detailed risk formulation depicts that the cloth mask fails in providing any significant protection as it reduces the infection risk as compared to the no-mask scenario by approximately 26% only as depicted by the risk comparison between no mask and cloth mask in Fig.13. The droplet filtration corresponding to cloth-mask has been depicted in Table 2.

| Table 2: Effect of cloth mask on ejected breathing droplet size | | | |
|---|---|---|---|
| Sr. No. | Size(μm) | Particle/second(cloth Mask) | Particle/second(No Mask) |
| 1 | 0.3-0.5 | 81 | 87 |
| 2 | 0.5-1 | 42 | 46 |
| 3 | 1-3 | 17 | 22 |
| 4 | 3 | 1 | 2 |

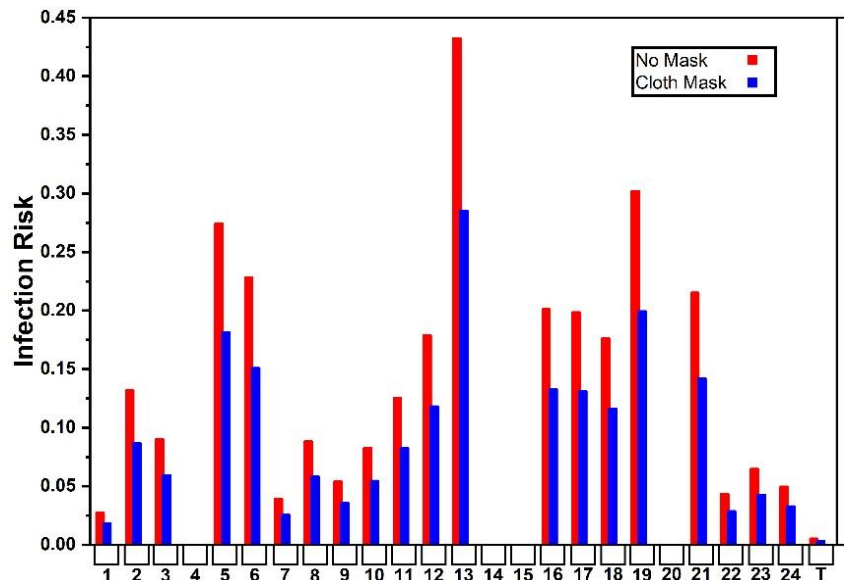

**Figure 13: Comparing airborne-infection risk for all the susceptible persons with no student wearing any mask, and students wearing cloth mask and surgical mask scenarios showing the contribution of each infected student as well(\*4,14,15,20 are themselves infected hence no droplet count has been shown).**

**III.A. Risk transmission associated with shields on desks ( considering the students are infected)**

Exploring the mask strategy proved to be unsuccessful in the previous discussion. Hence, a fairly ubiquitous practice in classrooms of having desk with shields installed on them has been investigated in this section[76]. Furthermore, there is also significant literature data suggesting the use of shield for infection risk minimization[28]. Droplets ejected during breathing, as evidenced by the diameter distribution enumerated in Table 1 remain in the aerosol range causing them to ascend due to thermal stratification caused by the body temperatures of the persons in the classroom, buoyancy and remain suspended in domain . Droplet dispersion plots in Figs. 14-15 where shields have been installed , depict that the shields are not able to completely entrap all the rising droplets albeit a significant percentage get trapped on them, thus pose a significant infection risk as evidenced by the risk comparison plot in Fig. 16 reducing the infection risk only by 50%.

Another form of menace posed by the shields is the fomite pathway of infection transmission. The shields, being attached with the benches are susceptible to touching by various persons. Furthermore, the shields are also composed of polycarbonate materials and the coronavirus can survive on such materials for quite a few days[77] thus further aggravating the precarious situation in absence of a proper sanitization practice. The risk of fomite infection in each shield has been evaluated using the Nicas and Best model[64] ( equations 3-5). Owing to the significantly less decay rate on the polycarbonate surfaces, Nicas and Best model[64] has been implemented using relevant literature data[64–66]. The plot in Fig. 17 depicts that the shields installed on the desk of the infected students pose a significant fomite risk.

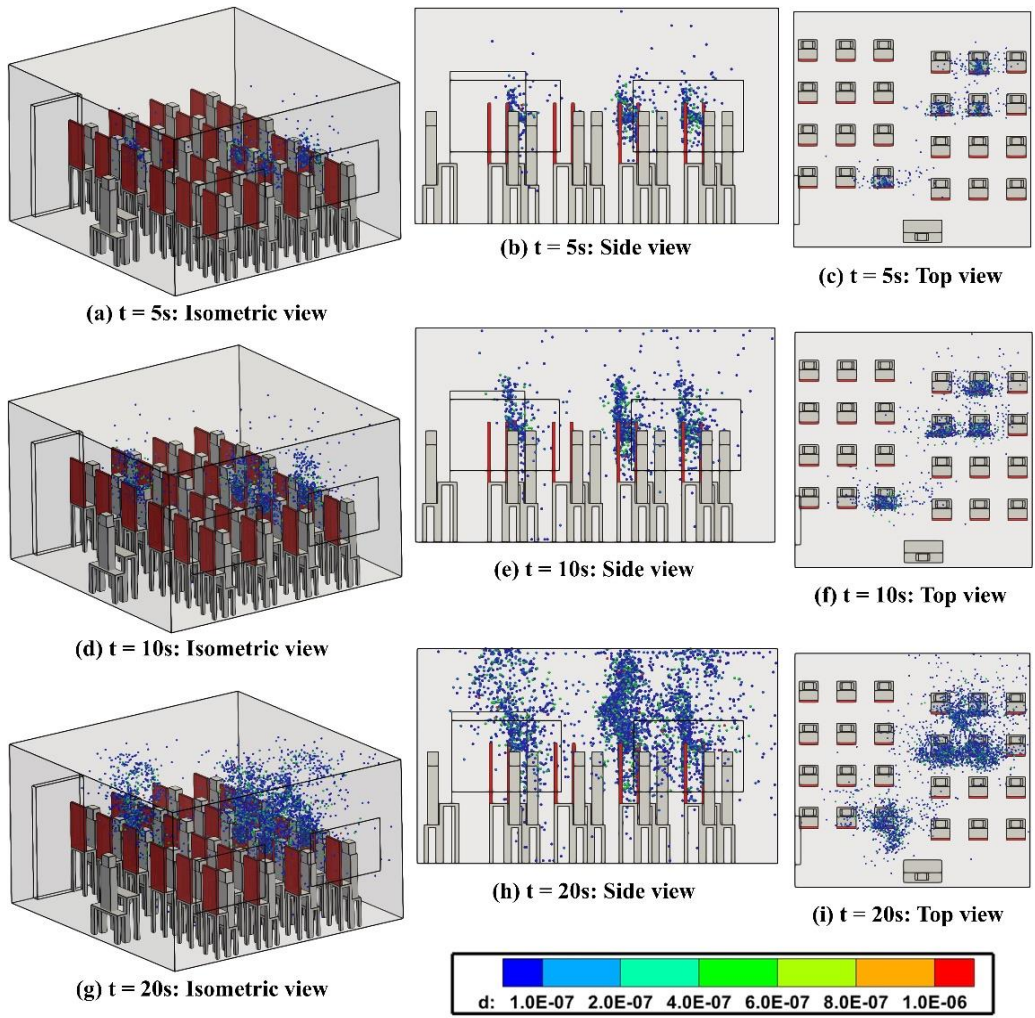

**Figure 14(a)-(i): Droplet dispersal in the domain with shields ( up to 20s)**

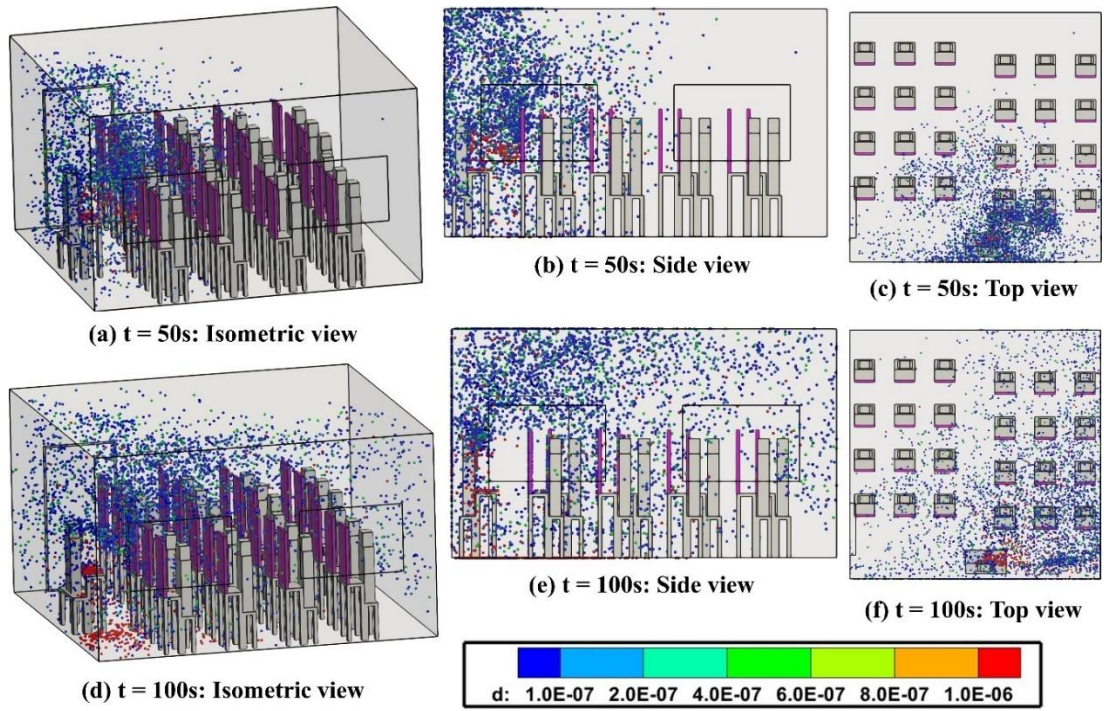

**Figure 15(a)-(f): Droplet dispersal in the domain with shields ( up to 100s)**

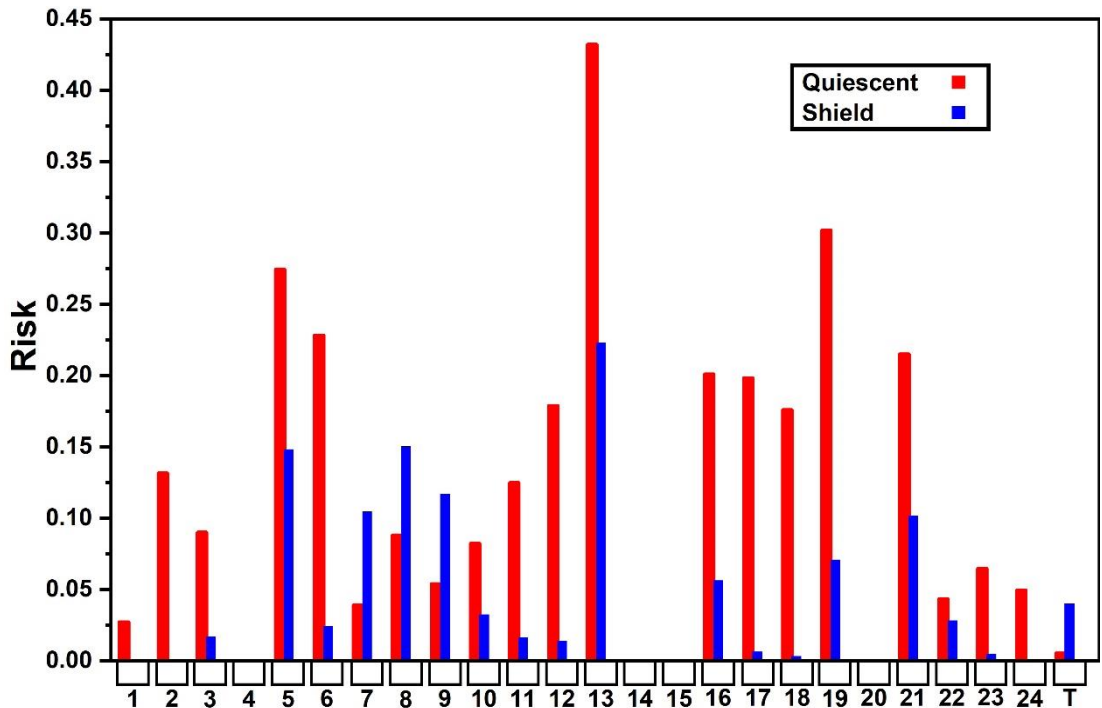

**Figure 16: Comparison of airborne-infection risk probability of susceptible persons between no-shield/screen and shield installed in classroom(*no bars are shown for 4,14,15,20 since they themselves are infected).**

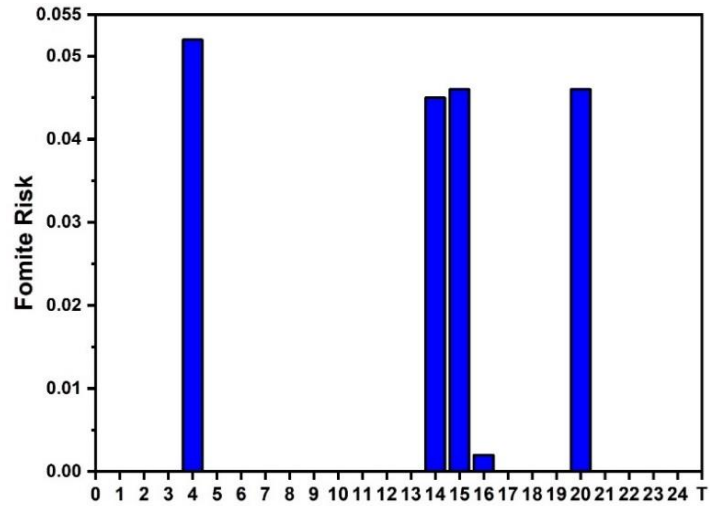

**Figure 17: Depicting the fomite risk of the shield of the students.**

**III.B. Risk transmission associated with shields on desks ( considering the teacher is infected)**

A significant percentage of droplets ejected during breathing and speaking[51,52] are either in sub-micron range or evaporate to become such. These droplets ascend in the domain due to lift and buoyancy force however after traversing certain distance they get entrapped in the downward loop of the convection loop in the domain generated due to the thermal stratification produced due to body temperatures of the persons in the classroom as evidenced by the droplet dispersion Fig. 18-19 and the velocity vector plots in Fig.20. Hence, after ascending over a certain distance they descend down to the student's breathing box height. During the descent, a significant percentage of droplets elude the shields and reach the vicinity of the susceptible students thus increasing the infection risk. The infection risk in the presence of shields has been portrayed in Fig.21 which shows in spite of installing shields a significant risk is sustained in the domain and for certain cases the infection probability reaches 100%.

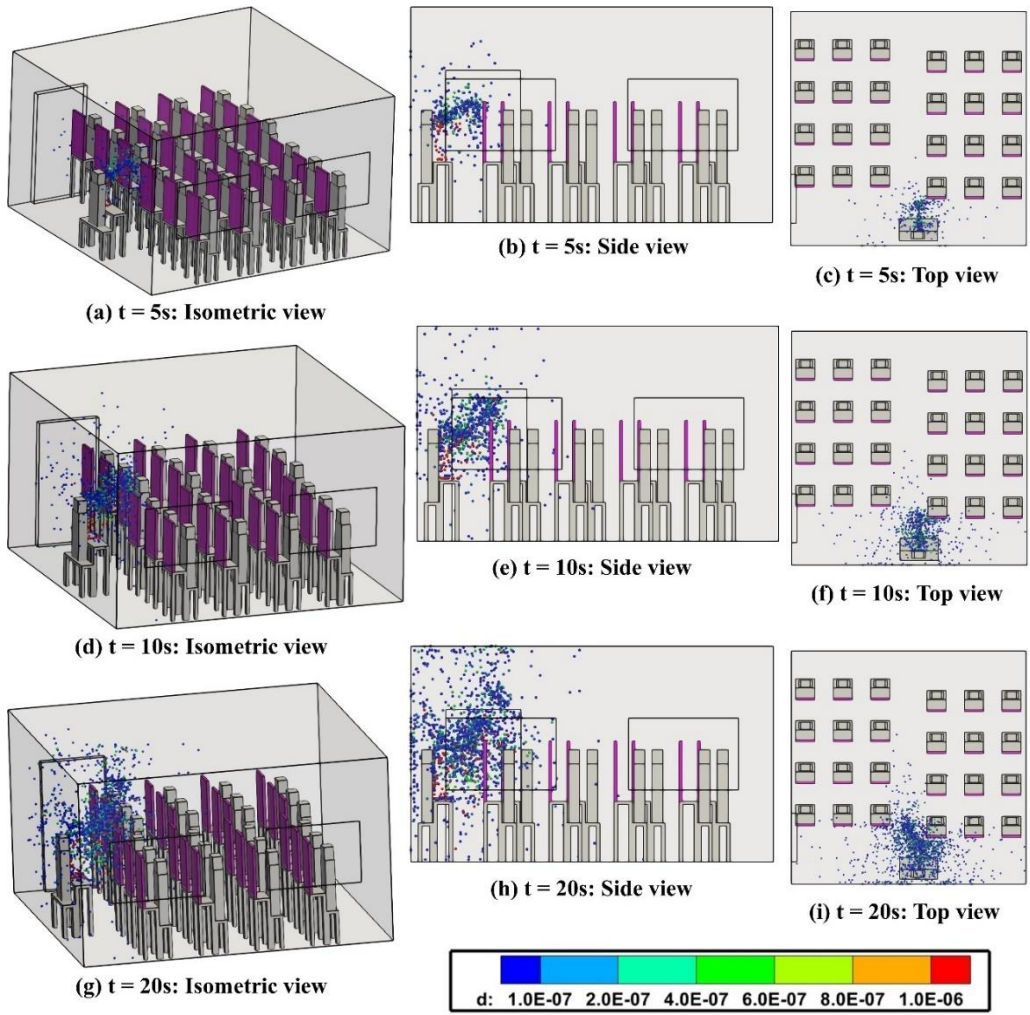

**Figure 18(a)-(i): Droplet dispersal in the domain with shields ( up to 20s)**

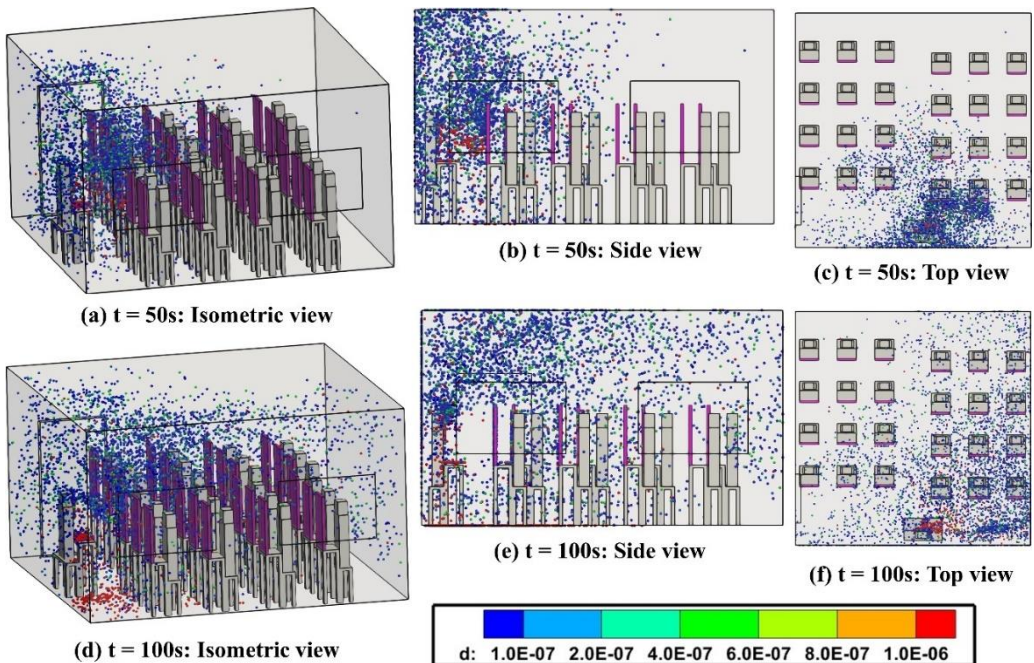

**Figure 19(a)-(f): Droplet dispersal in the domain with shields ( up to 100s)**

**Submitted to Arxiv**

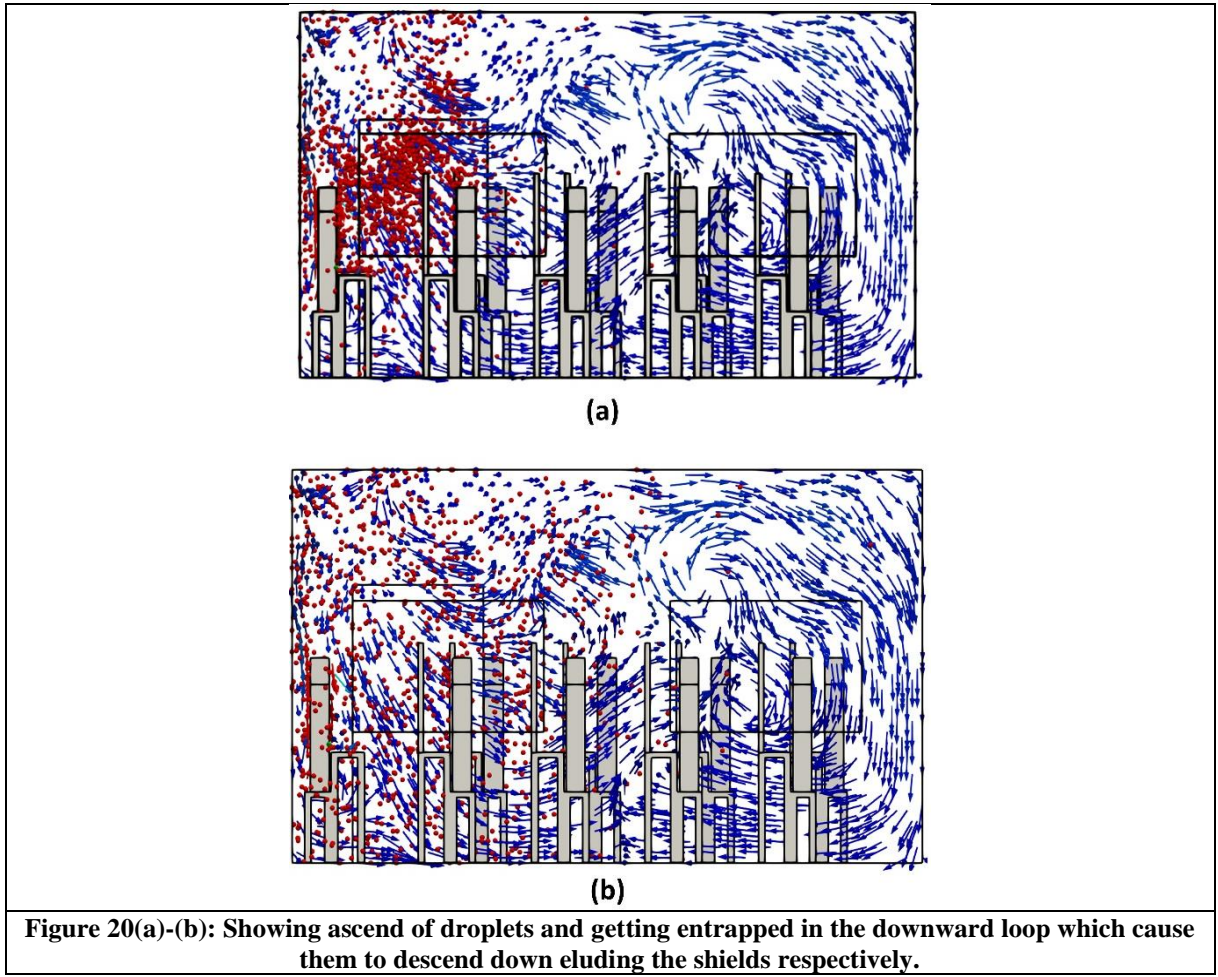

**Figure 20(a)-(b):** Showing ascend of droplets and getting entrapped in the downward loop which cause them to descend down eluding the shields respectively.

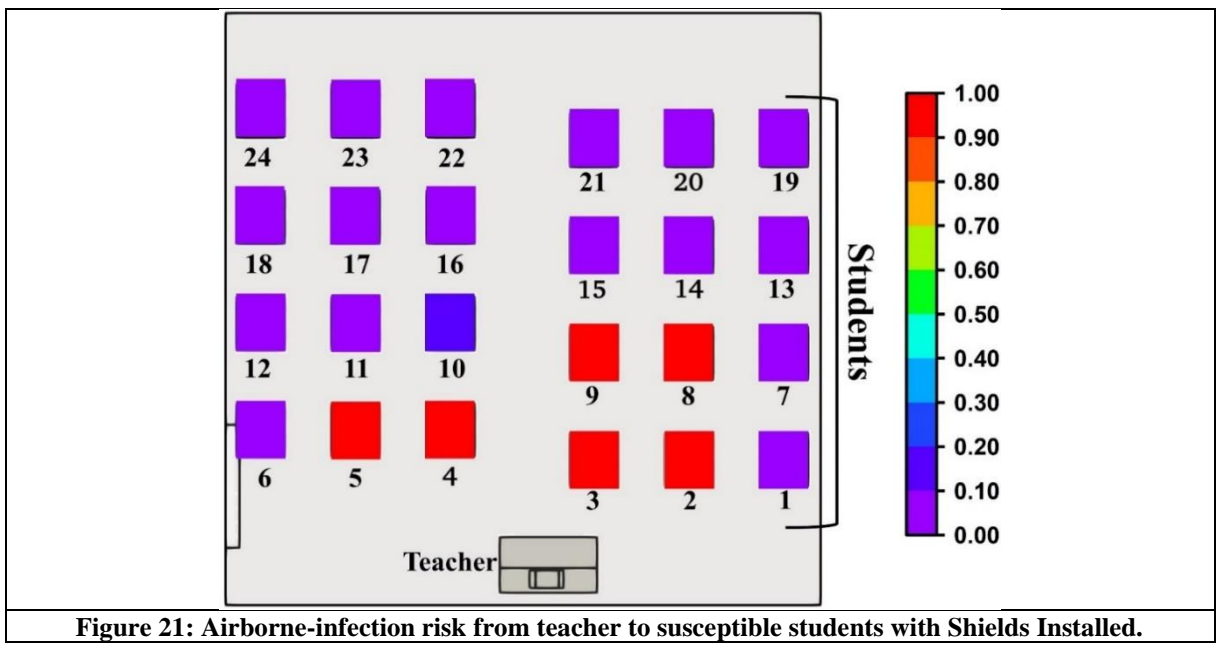

**Figure 21:** Airborne-infection risk from teacher to susceptible students with Shields Installed.

## IV.A. Risk reduction through polycarbonate Screens (considering the students are infected)

It can be conclusively argued from previous discussions that the injected droplets ascend in domain upon injection due to convection, lift and buoyancy. Furthermore, due to their size remaining in the aerosol range[51,52] as depicted in Table 1 they remain suspended in the domain at or above the breathing boxes' height for significant period. The droplet dispersion juxtaposed with velocity vector in Figs.4-5 upholds this argument. Hence, an effective way for airborne risk reduction would be entrapping and arresting the movement in droplets during their upward dispersal. Henceforth, a very low-cost sustainable solution in the form of transparent polycarbonate screens suspended from the ceiling above the head height of the sitting person, is proposed to accomplish the prevention of upward dispersal of droplet as depicted in Fig. 22. The droplet dispersals in the presence of screens, in Figs.23-24 depict that a significant percentage of injected droplets get entrapped on the screen thus decreasing the airborne-infection risk significantly. A comparison of risk in absence of screens and that in the presence of the screens, depicted in Fig 25 confirms the fact that there is a significant decrease in infection risk, the reduction in risk of the most susceptible person being 92% from the no-screen scenario (quiescent ventilation discussed above is the no-screen scenario). Furthermore, another important aspect that needs to be considered is that the screens are more effective in providing protection against the cloth mask. Thus demonstrating the effectiveness and the need of installing the screens in providing protection against risk propagation. Hence, it can be safely concluded that the screens not only reduce the infection risk but also eradicates the significant epidemiological implication associated with it by significantly reducing the droplet count that reaches the susceptible students.

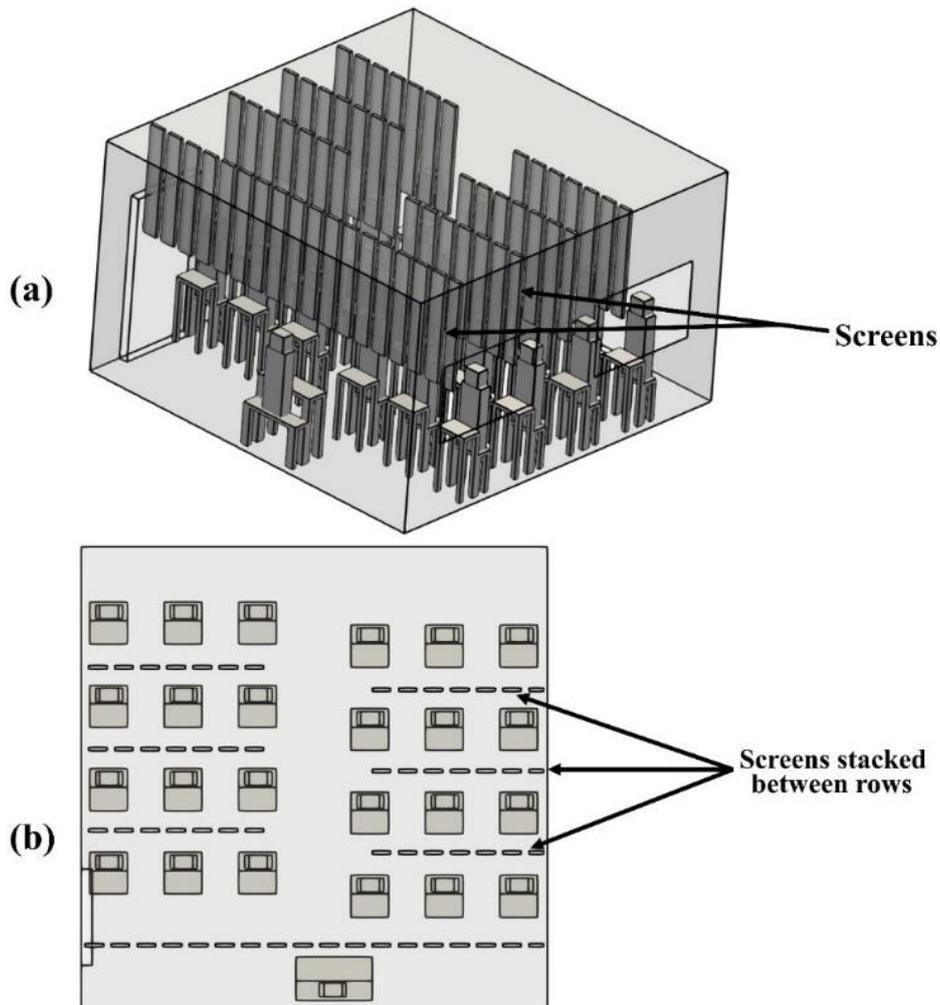

**Figure 22 (a)-(b): Depicting the suspended screens from the ceiling**

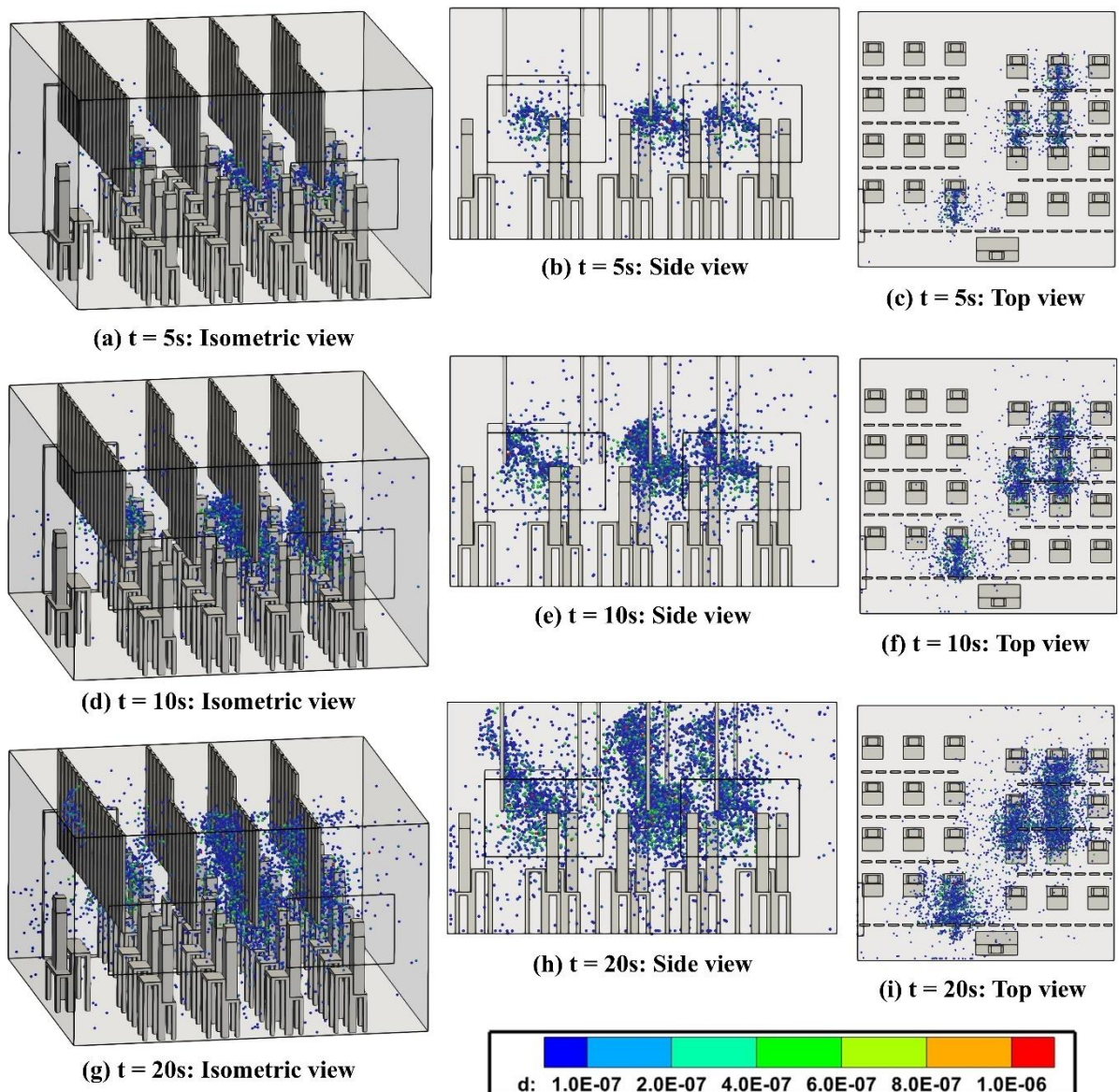

**Figure 23(a)-(i): Droplet dispersal in the domain with screens showing a significant percentage of droplets get arrested by the screens ( up to 20s)**

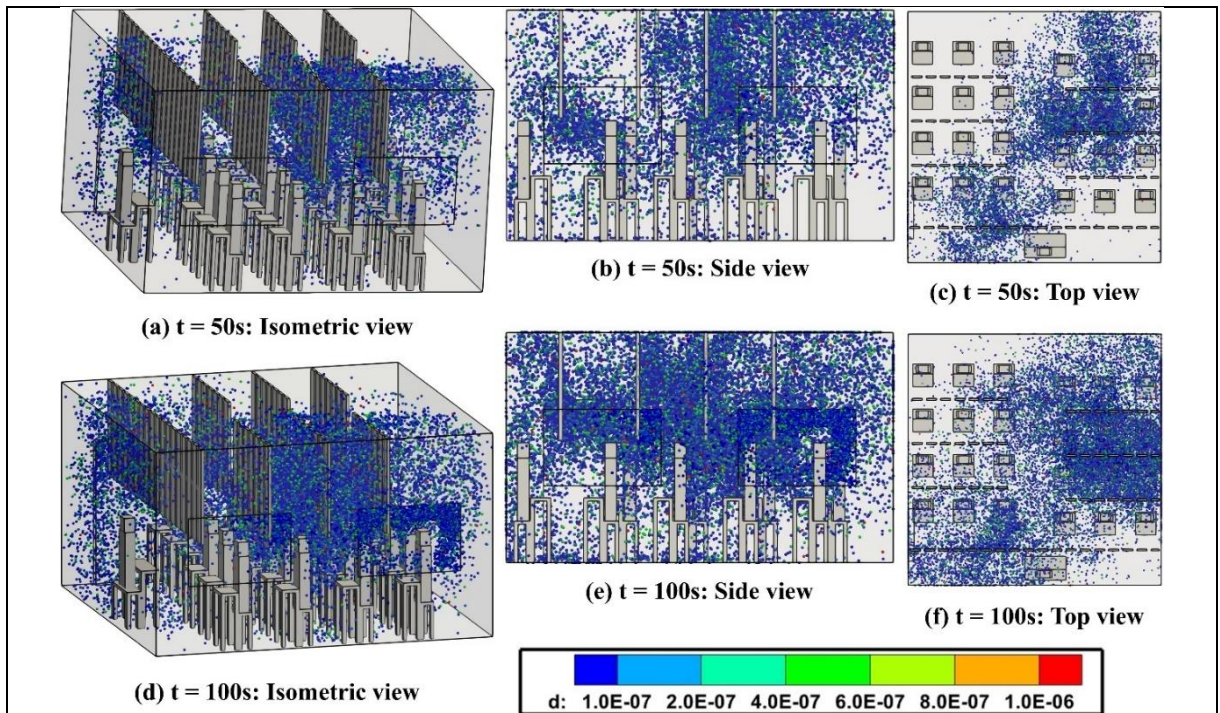

**Figure 24(a)-(f):** Droplet dispersal in the domain with screens showing a significant percentage of droplets get arrested by the screens (up to 100s)

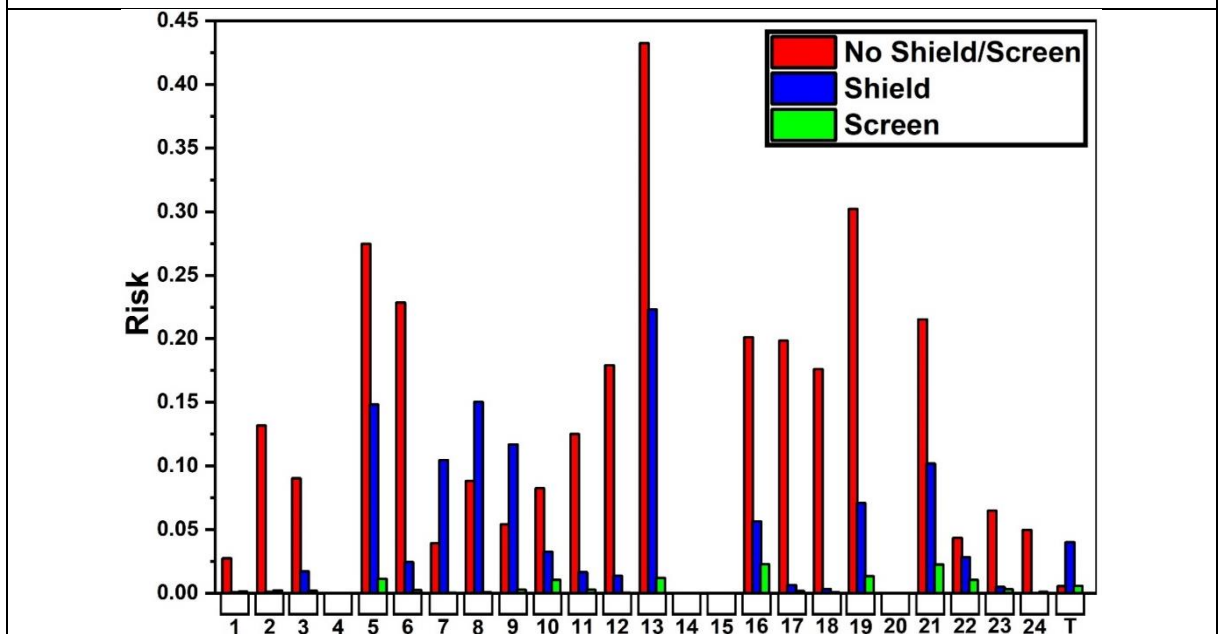

**Figure 25:** Comparison of airborne-infection risk probability of susceptible persons between no-shield/screen, shield and screen installed in classroom(*no bars are shown for 4,14,15,20 since they themselves are infected).

**Advantages over shields on desk scenario.**

Solutions in the form of shield installed on shields are applicable for situations pertaining to larger droplet size mainly emanated in respiratory activities like coughing[58] and sneezing as the droplets upon ejection tend to descend down and hence their motion is arrested by the shield which ultimately prevents the lateral spreading leading to risk eradication. Infected persons showing symptoms like coughing, sneezing can easily be identified and isolated, however asymptomatic persons ejecting sub-micron droplets into domain during breathing are difficult to identify. Droplets ejected during breathing are much smaller as compared to coughing or sneezing[51,52], as evidenced by the diameter distribution enumerated in Table 1 and hence much lighter which ultimately causes them to remain suspended and ascend in domain due to reasons discussed previously. The failure of inhibition of the ascent of the droplets by the shields is what renders them ineffective against these kind of droplets. On the other hand the screens suspended from the ceiling can arrest the upward ascent of the droplets and stop the lateral spread of these virus containing droplets thus significantly decreasing the infection risk. Figure 25 demonstrates that the screen is 2.16 times more efficacious than shield in terms of risk reduction. Furthermore, the screens due to their strategic location enjoy an inherent advantage over shields in the aspect of fomite mode of infection. Students will not be predisposed to touch the screens as the latter are stacked down from the ceiling above the students' seating height and due to their absence in the immediate vicinity of any particular seated student, as can be observed from Fig. 22.

**IV.B. Risk reduction associated with screens (considering the teacher is infected)**

Similar to the scenario of infected students, the transparent polycarbonate screens provide substantial protection by entrapping the droplets ejected into the domain by the infected teacher during their ascent, thus preventing a significant percentage them from reaching the vicinity of students' breathing box. The effect of screen on dispersal of droplets emanating from the infected teacher depicted in the Figs.26-27 explains the above-mentioned phenomenon. The airborne-risk reduction of the most susceptible person by 95 % from the quiescent scenario through the installation of screens can be understood from Figs.28. The reduction of infection-risk through the fomite route is 100% as there is insignificant fomite-infection risk in the case of screens. Similar, to the screens scenario for the students the shields provide protection against the epidemiological implications associated with the droplets by significantly reducing the droplet count that reaches the students.

**Advantage over shields installed on desks**

Just like the scenario for infected students screens provide significant advantage over the shields for the infected teacher scenario as well. In the case of the shields the infection probability reaches 100% since the droplets elude the shields and reach the students thus posing a significant risk to the students. On the other hand the screens suspended from the ceiling arrest the motion the droplets and this ultimately reduces the infection risk significantly as depicted in Fig. 28.

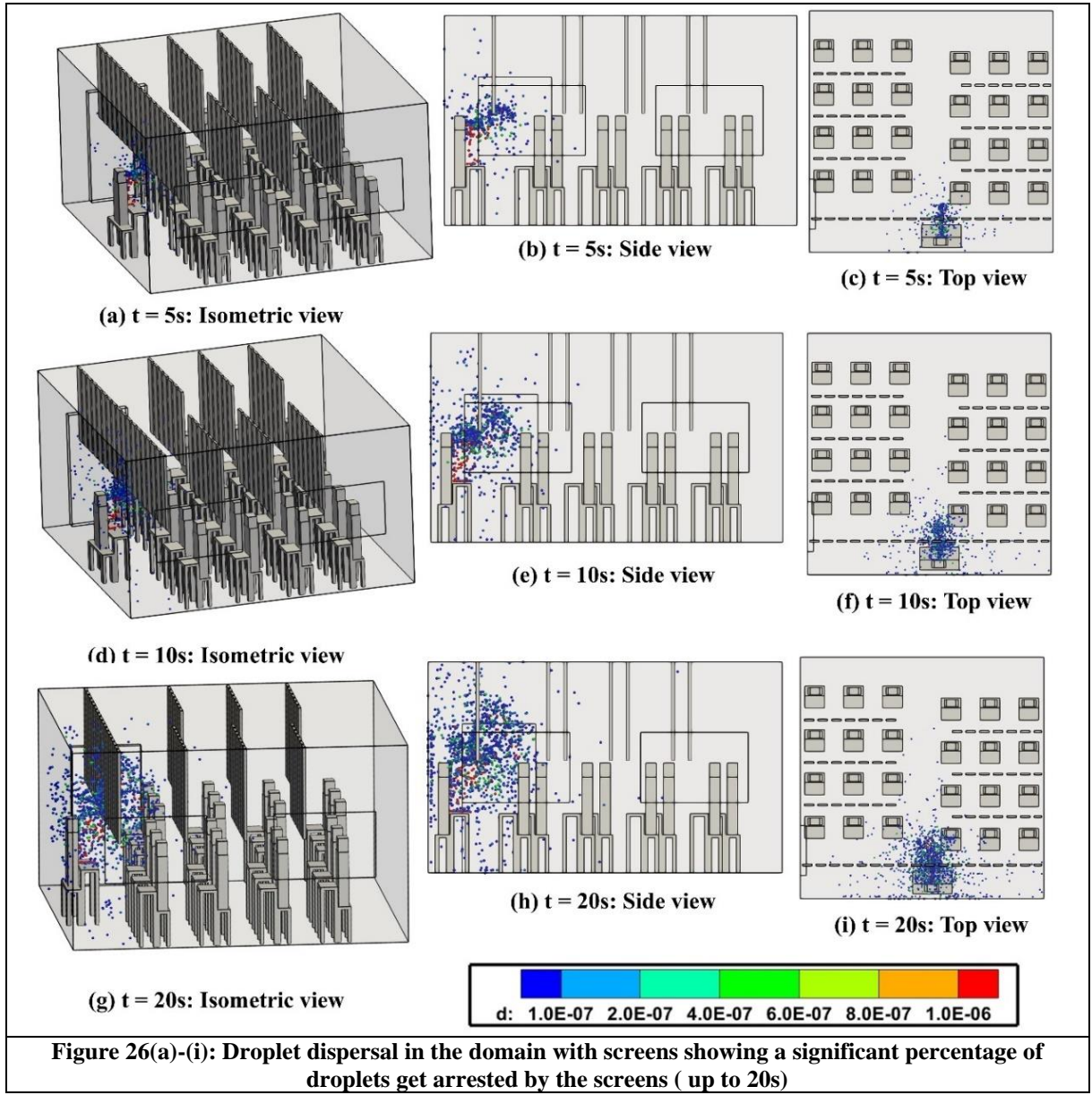

Figure 26(a)-(i): Droplet dispersal in the domain with screens showing a significant percentage of droplets get arrested by the screens ( up to 20s)

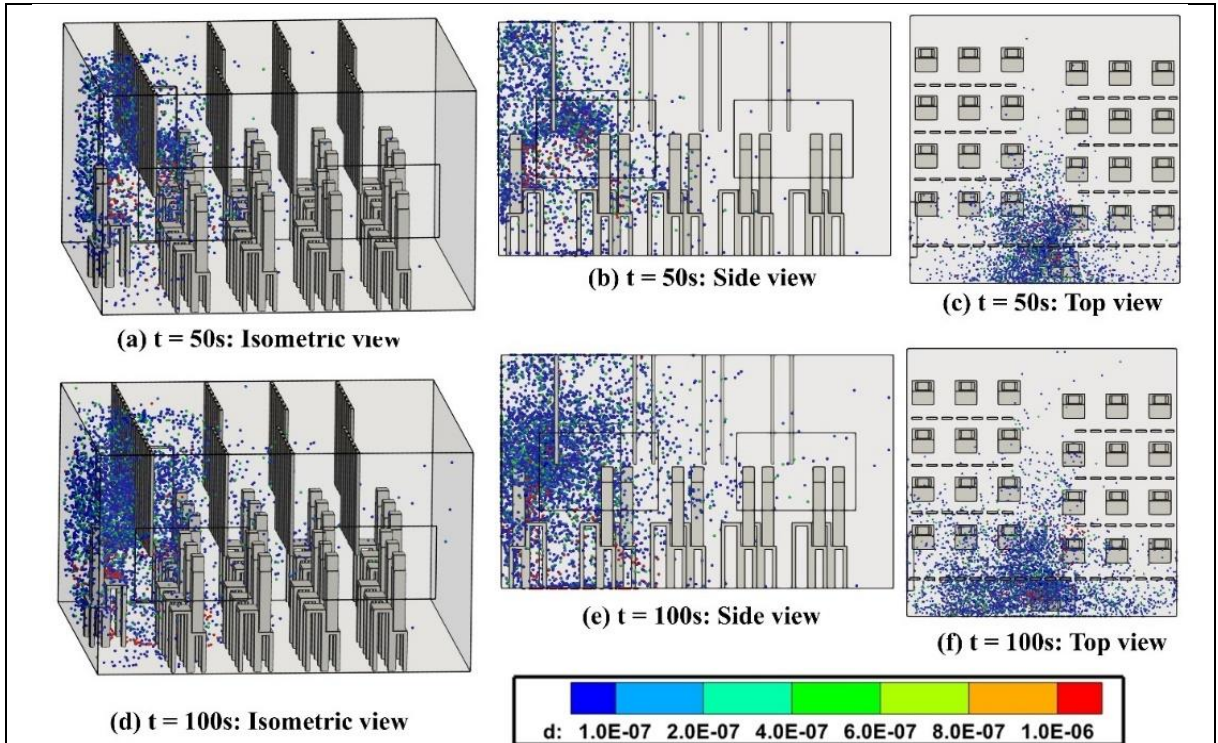

**Figure 27(a)-(f): Droplet dispersal in the domain with screens showing a significant percentage of droplets get arrested by the screens (up to 100s)**

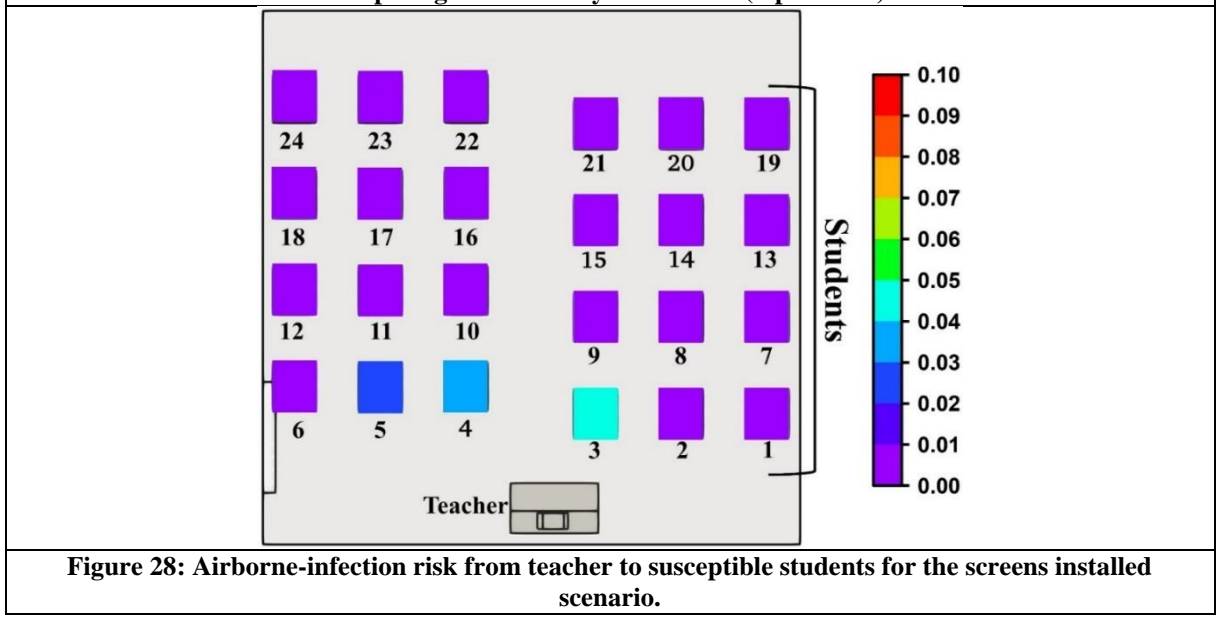

**Figure 28: Airborne-infection risk from teacher to susceptible students for the screens installed scenario.**

### 3.2. Predicting the Class Time

The maximum infection risk reduction in the case of the teacher being infected is more than that of the four students being infected through screens as can understood from Figs 20 and 14. Furthermore, the viral load injected into the domain when four students are infected is way more than when the teacher is infected. In the no-mask, no-shield/screen it has been observed that the maximum infection-risk probability reaches 50% within a period of 108s as depicted in Fig.2 of Appendix II. As can be understood from the installation of screens creates a very safe situation and ensures that the class can be conducted for an extended period. Hence, the safe time for which the class can be conducted considering the students are infected when the transparent polycarbonate screens are installed has been predicted. A linear fit of the maximum infection risk variation with time over the exposure period of 100 seconds was obtained as depicted in Fig.3 in Appendix II. The fitted graph was extended till the infection risk probability reaches 50%, since continuing class beyond this period becomes very risky. This time-period comes out to be approximately 3312 s. or 55 mins. To check the authenticity of the fit the simulation was executed for 1000s and the data beyond 100s up to 1000s shows very less deviation as has been depicted in Fig.4 in the appendix section, thus proving the authenticity of the obtained fit. A similar linear extrapolation was conducted for the scenario of shields for depicting the efficacy of shields in extending class time in Fig.5. A safe-class time of approximately 225 seconds was determined thus proving that the shields are practically useless in providing protection against virion containing droplets ejected during breathing. Table 2 below provides the predicted safe-class time for different scenarios. The cloth-mask scenario has not been depicted below, since the cloth-mask does not provide significant risk reduction from the quiescent scenario.

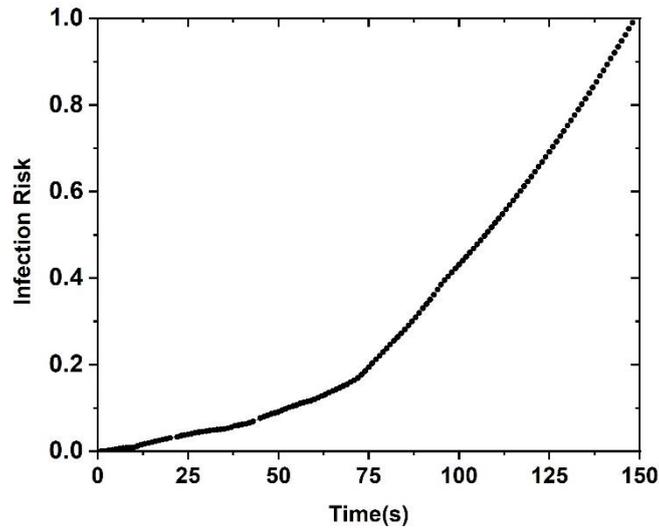

**Figure 25: Showing maximum airborne-infection risk probability reaches 50% within 108s for the quiescent scenario.**

| Table 2: Safe Class Time | | |
|---|---|---|
| **Sr. No.** | **Scenario** | **Safe Class Time(s)** |
| 1 | Quiescent ( devoid of Screen/Shield) | 108 |
| 2 | Shield | 225 |
| 3 | Screen | 3312 |

4.Conclusions

In this study the droplet dispersion and the associated epidemiological implications has been investigated in a classroom constituting 24 students and one teacher. The infection risk associated with two distinct scenarios, when students are infected and when the teacher is infected has been reported for a hot dry quiescent scenario. Speaking while wearing a face mask causes vocal strain, vocal fatigue, loss of breath and requires talking at an elevated level. This study pertains to all socio-economic conditions and in the developing countries classrooms of the educational institutions are devoid sound-relaying systems. Hence to ensure an extended class with no discomfort to the teacher and make the sound audible to all the students, it has been assumed that the teacher is not wearing any mask. The airborne infection risk in quiescent scenario cumulates to a very high value, 30% for the scenario of infected students and 100% for the scenario of infected teacher. The magnitude of basic reproduction number($R_0$) associated with an infected student or an infected teacher in a quiescent medium is 6 and 7 respectively. Another important risk pertaining to the quiescent scenario that was explored is the epidemiological implications of the droplet size. The droplet size ejected during the exhalation activity has a very deposition probability in lower parts of respiratory tract that is in the alveolar and bronchial regions, which can lead to long term lungs damage. Also, it was found out that a significant percentage can deposit in the olfactory epithelium. This not only leads to olfactory dysfunction and anosmia but also opens up potential damage to central nervous system and ultimately the brain. The virions can affect the brain through the olfactory nerves. Thus, besides posing a significant infection risk these droplets have the potential causing long term lung and neurological damages. Hence, it is pertinent complete protection against these droplets must be devised.

The fairly ubiquitous cloth mask proves ineffective in providing any substantial protection just like another commonly used strategy namely shields installed on desks. The masks reduced the maximum infection probability by approximately 26 % only. The shields were able to bring only a 50% risk reduction in the scenario of infected students besides themselves becoming hotspots for fomite infection. In the case of infected teacher, the ejected small droplets ascended due to buoyancy and lift and remain suspended in the domain. During this process they eluded the shields an reached the vicinity of students causing the infection risk to reach 100% for certain students.

A novel solution in the form of screens suspended from the domain has been proposed to entrap the droplets and reduce the infection risk. The small sized droplets ejected during breathing ascend in the domain due to buoyancy, lift and the natural circulation loop in the domain created due to the thermal stratification generated in the domain due to body temperatures of the person. The screens entrap a significant percentage of droplets during their ascent and prevent them from reaching the vicinity of persons, thus reducing the infection risk significantly. The screens have proven to be effective in both the scenarios of the teacher and student being infected reducing the infection risk by 95% and 92% respectively for the scenario of infected teacher and student respectively. A maximum class time that ensures low-risk for the susceptible persons has been predicted. The infection-risk over the exposure time-period of 100s has been fitted and a good fit has been obtained which agrees with data beyond 100s up to 1000s. A maximum class duration of 3312s or 55 mins has been proposed. The maximum safe-class time of 108s for the Quiescent ( no-shield/screen) was determined and the same for the shield scenario was 225s.

**Acknowledgement**

We want to express our gratitude to the High-Performance Computing Cluster at Jadavpur University's Technological Bhavan for helping us to complete the simulations in a timely manner.

**Appendix Section**

**I. Risk heat-map associated with each student**

As discussed in the main manuscript the heat-map pertaining to each infected student has been presented in Fig.1. The risk pertaining to the individual persons from themselves has been depicted by black boxes since an infected person cannot infect oneself. The vertical axis represents the infected persons and the horizontal axis represented the susceptible persons (T- represents the teacher). In the risk-matrix depicted below each cell represents $R_{i-j}$ of each individual combination( including the teacher). $R_{i-j}$ obtained from this matrix is substituted in equation 3 to obtain the total infection risk posed by the $i^{th}$ infected person.

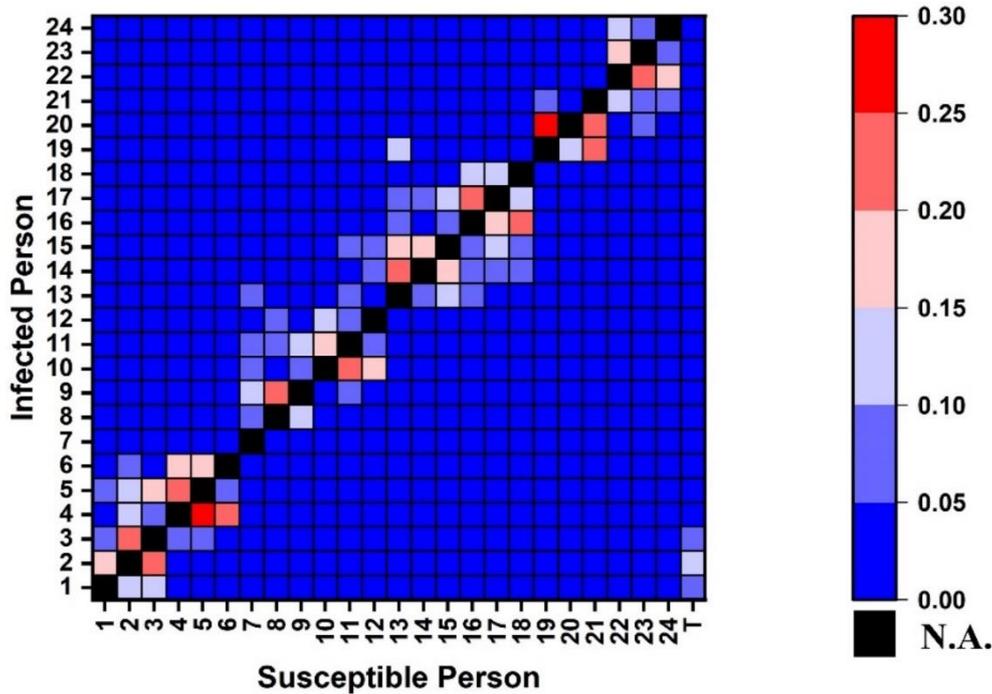

**Figure 1: Risk heat map representing the risk posed by each infected person to the susceptible persons(* the risk pertaining to the individual persons from themselves has been depicted by black boxes since an infected person cannot infect oneself., T- signifies teacher)**

## II. Prediction of the Safe-Class Time

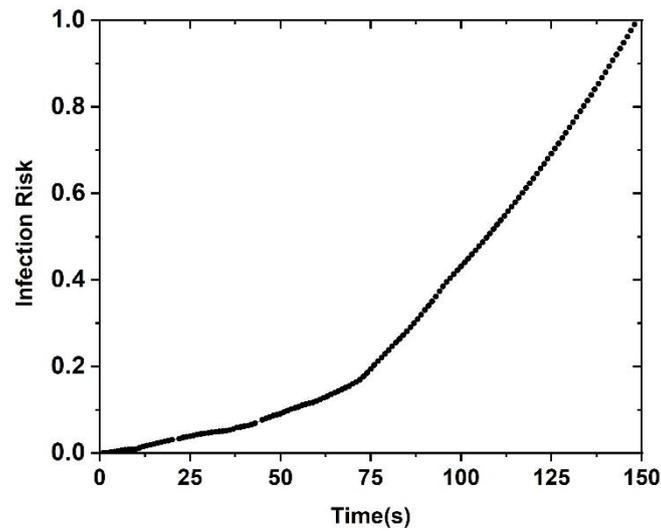

**Figure 2: Showing maximum airborne-infection risk probability reaches 50% within 108s for the quiescent scenario.**

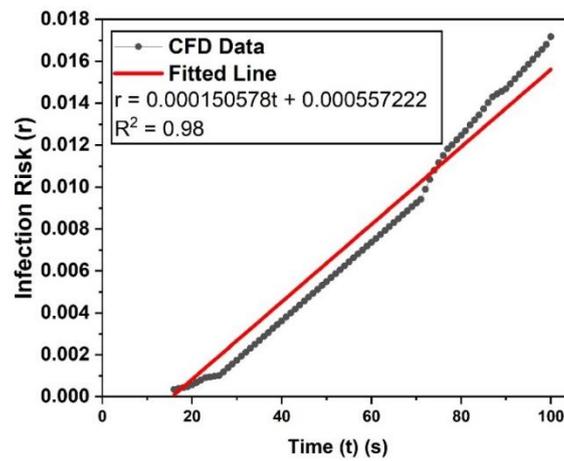

**Figure 3: Plotting the fit of variation of maximum airborne-infection risk with time for screens.**

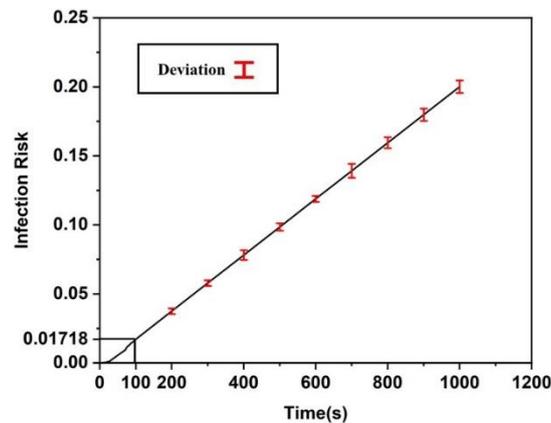

**Figure 4: Showing deviation of actual maximum airborne-infection risk against fitted data for screens.**

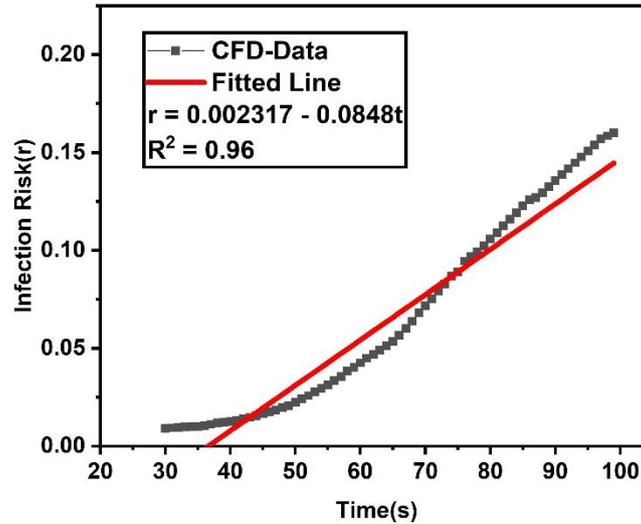

**Figure 5: Plotting the fit of variation of maximum airborne-infection risk with time for shield.**

Figure 2 depicts the risk variation with time and maximum safe-class time in the Quiescent (no Shield/Screen) to be 108s. Figure 3 depicts the linear-fit of the risk variation with time for the screen scenario over the exposure time of 100s. Figure 4 depicts the deviation of the actual infection risk from the obtained linear fit from the infection risk over the exposure period of 100sec. The deviation remains within 10% thus the fit can be used for predicting the safe class duration. Figure 5 depicts the linear-fit of the risk variation with time for the shield scenario over the exposure time of 100s.

### III. Grid-Independence and Time-Independent Study

**Grid-Independence Study**

A 3D structured hexahedral-dominant mesh is used to discretize the computational domain in our model. There are approximately $1.17 \times 10^6$ cells total in the mesh. A sufficient amount of mesh refinement is employed close to any person's mouth, where the respiratory droplets are injected, in order to correctly capture the droplet motion. This refinement is applied around benches, desks, the students, and the teacher. The transition from the finely refined mesh close to the mouth to the background mesh was also smooth and gradual. A thorough grid independence investigation was conducted before choosing the aforementioned mesh. Coarse ($9.37 \times 10^5$ cells), medium ($1.17 \times 10^6$ cells), and fine mesh sizes were chosen ($1.4 \times 10^6$ cells). To perform a detailed grid independence study encompassing both the Eulerian and Lagrangian fields, the maximum risk of infection over exposure period of classroom time of 100s is taken as the parameter for comparing the results of the 3 different meshes. It can be observed from Fig.6 that the results of the coarse mesh differ significantly from those of the medium and fine meshes, while the results of the medium and fine meshes differ only marginally. Because the medium and fine meshes produce findings that are practically grid independent, we chose the medium mesh for our investigation.

**Time-independence Study**

Table 1 below depicts the time-independence study in detail. Three different time-step sizes of $10^{-3}$, $10^{-4}$, $10^{-5}$ were selected and the corresponding magnitude of the risk of infection is shown for these time steps. Results indicate that there is significant variation in results between time step sizes of $10^{-3}$ and $10^{-4}$ – with a 3% change in the magnitude of risk of infection however there is no significant difference in results between time-step sizes of $10^{-4}$ and $10^{-5}$ – with a percentage change of less than 0.1% only. The results are practically time-independent. Hence, the time-step size of $10^{-4}$ has been selected for this study.

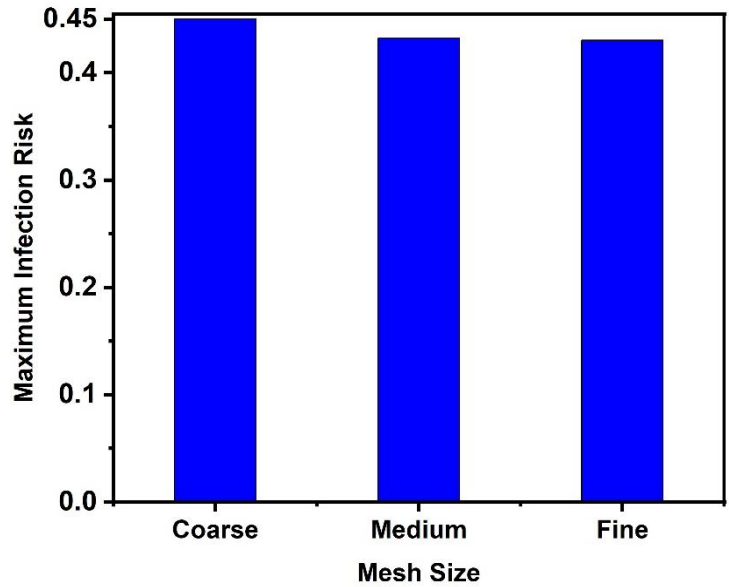

**Figure 6: Comparison of the maximum risk of infection of different mesh sizes .**

Table 1: Showing time-independence data.

| Time step size | Magnitude of maximum infection risk | Percentage change from previous time step value |
|---|---|---|
| $10^{-3}$ | 0.4458 | |
| $10^{-4}$ | 0.43239 | 3% |
| $10^{-5}$ | 0.4319 | 0.1% |

# Submitted to Arxiv

**Submitted to Arxiv**